\documentclass{aa}  

\usepackage{graphicx}
\usepackage{txfonts}
\usepackage{xcolor}
\usepackage{subfig}
\usepackage{makecell}
\begin{document}

   \title{High resolution optical spectroscopy of the $\mathrm{N_2}$-rich comet C/2016 R2 (PanSTARRS)}

   \subtitle{}

   \author{C. Opitom
          \inst{1}
          \and
          D. Hutsem\'ekers\inst{2}
          \and
          E. Jehin\inst{2}
          \and
          P. Rousselot\inst{3}
          \and
          F. J. Pozuelos \inst{2}        
          \and
          J. Manfroid\inst{2}
          \and 
          Y. Moulane \inst{2,4}
          \and
          M. Gillon \inst{2}
          \and
          Z. Benkhaldoun \inst{4}
          }

   \institute{ESO (European Southern Observatory) - Alonso de Cordova 3107, Vitacura, Santiago Chile \\
         \email{copitom@eso.org}
         \and STAR Institute - University of Liège, All\'ee du 6 Ao\^ut 19c, B-4000 Liège, Belgium
         \and Institut UTINAM-UMR CNRS 6213, Observatoire des Sciences de l’Univers THETA, University of Franche-Comt\'e, BP 1615, F-25010 Besan\c{c}on Cedex, France
         \and Oukaimeden Observatory, High Energy Physics and Astrophysics Laboratory, Cadi Ayyad University, Marrakech, Morocco\\
             }

   \date{}

 
  \abstract
   {Early observations of comet C/2016 R2 (PanSTARRS) have shown that the composition of this comet is very peculiar. Radio observations have revealed a CO-rich and HCN-poor comet and an optical coma dominated by strong emission bands of $\mathrm{CO^+}$ and, more surprisingly, $\mathrm{N_2^+}$.}
   {The strong detection of $\mathrm{N_2^+}$ in the coma of C/2016 R2 provided an ideal opportunity to measure the $\mathrm{^{14}N/^{15}N}$ isotopic ratio directly from $\mathrm{N_2^+}$ for the first time, and to estimate the $\mathrm{N_2/CO}$ ratio, which is an important diagnostic to constrain formation models of planetesimals, in addition to the more general study of coma composition.}
   {We obtained high resolution spectra of the comet in February when it was at 2.8 au from the Sun. We used the UVES spectrograph of the ESO VLT, complemented with narrow-band images obtained with the TRAPPIST telescopes.}
   {We detect strong emissions from the ions $\mathrm{N_2^+}$ and $\mathrm{CO^+}$, but also $\mathrm{CO_2^+}$, emission from the CH radical, and much fainter emissions of the CN, $\mathrm{C_2}$, and $\mathrm{C_3}$ radicals which were not detected in previous observations of this comet. We do not detect OH or $\mathrm{H_2O^+}$, and derive an upper limit of the $\mathrm{H_2O^+/CO^+}$ ratio of 0.4, implying that the comet has a low water abundance. We measure a $\mathrm{N_2^+/CO^+}$ ratio of $0.06\pm0.01$. The non-detection of $\mathrm{NH_2}$ indicates that most of the nitrogen content of the comet lies within $\mathrm{N_2}$. Together with the high $\mathrm{N_2^+/CO^+}$ ratio, this could indicate a low formation temperature of the comet, or that the comet is a fragment of a large differentiated Kuiper Belt object. The $\mathrm{CO_2^+/CO^+}$ ratio is $1.1\pm0.3$. We do not detect $\mathrm{^{14}N^{15}N^+}$ lines, and can only put a lower limit on the $\mathrm{^{14}N/^{15}N}$ ratio measured from $\mathrm{N_2^+}$ of about 100, compatible with measurements of the same isotopic ratio for $\mathrm{NH_2}$ and CN in other comets. Finally, in addition to the [OI] and [CI] forbidden lines, we detect for the first time the forbidden nitrogen lines [NI] doublet at 519.79 and 520.04 nm in the coma of a comet.}
   {}


   \maketitle
%

\section{Introduction}

Comets are among the most pristine bodies of the Solar System. They have undergone little alteration since their formation and their nucleus preserves invaluable clues about the evolution of volatile material within the early Solar System at the time of planet formation. Cometary material reveals a diversity of formation processes dating from the early Solar System. In particular, isotopic abundance ratios measured in cometary material are very sensitive to local physico-chemical conditions through fractionation processes.

On 2016 September 7, the PanSTARRS telescope discovered the long period comet C/2016 R2 (PanSTARRS), hereafter R2 \citep{Weryk2016}. With a period of about 20000 years, and a semi-major axis of 735 au, R2 is a returning object coming from the Oort cloud, meaning that this was not its first passage close to the Sun \citep{levison1996}. The comet developed a coma at large heliocentric distances (r$\sim$6 au), and as it was getting closer to the Sun, in December 2017 (at about 3.0 au), it started to display a very unusual coma morphology in optical images, with a lot of structures changing rapidly. This was attributed to ions dominating the emission of the coma and emitting mainly at blue optical wavelengths. Because of its peculiarity, R2 has been the target of numerous telescopes around the world, and at various wavelengths. Radio observations taken near the end of December 2017 revealed a very CO-rich comet with a low abundance of HCN \citep{Wierzchos2017}. Optical spectra obtained the same month with the 2.7m telescope of the McDonald Observatory showed that its spectrum was largely dominated by the emission bands of $\mathrm{CO^+}$ and more surprisingly of $\mathrm{N_2^+}$ \citep{Cochran2018}, confirming that R2 is a very unusual and interesting object. 

The detection of $\mathrm{N_2}$ in comets has been a matter of debate for decades. The N$_2$ molecule itself cannot be detected in cometary spectra in the optical range, but N$_2^+$ can be observed in this range thanks to the bands of the first negative group (B$^2\Sigma_u^+$--X$^2\Sigma_g^+$), the (0,0) bandhead being at 391.4~nm. Before the apparition of R2 only very few detections of N$_2^+$ emission lines had been reported from ground based facilities. These concern very few comets, mainly: C/1908 R1 (Morehouse) \citep{Pluvinel1911}, C/1961 R1 (Humason) \citep{Greenstein1962},1P/Halley \citep{Wyckoff1989,Lutz1993}, C/1987 P1 (Bradfield) \citep{Lutz1993}, 29P/Schwassmann–Wachmann 1 \citep{Korsun2008,ivanova2018}, and C/2002 VQ94 (LINEAR) \citep{Korsun2008,Korsun2014}. From these detections only the last two are based on spectra with both good signal-to-noise ratio and good spectral resolution. It should also be pointed out that some spectra might have been contaminated by telluric N$_2^+$ emission lines. The first in situ detection of N$_2$ in a comet was done in 67P’s coma by the ROSINA mass spectrometer on-board the Rosetta spacecraft (Rubin et al. 2015) (with a N$_2$/CO ratio of 5.7$\pm$0.66$\times10^{-3}$).

The $\mathrm{N_2}/ \mathrm{CO}$ ratio provides clues about the formation temperature of comets. Following the first failed attempts to detect $\mathrm{N_2^+}$, several models were developed to explain the  $\mathrm{N_2}$ deficit in the coma of comets. \cite{Owen1995}, considering amorphous ice trapping $\mathrm{CO}$ and $\mathrm{N_2}$ gases at temperatures of about 50K, have shown that the $\mathrm{N_2}/ \mathrm{CO}$ abundance ratio in cometary ice would be of the order of 0.06. \cite{Iro2003} proposed a different model to account for low $\mathrm{N_2}/ \mathrm{CO}$ ratio considering the trapping of volatiles in clathrates, $\mathrm{CO}$ being trapped much more easily than $\mathrm{N_2}$. 
\cite{Mousis2012} considered that cometesimals were formed by the agglomeration of clathrate hydrates, together with other ices. Their model predicts a low abundance of molecular nitrogen in comets as a consequence of the formation of planetesimals at very low temperature (in the 22-47K range) allowing formation of pure $\mathrm{N_2}$ condensate. Subsequent evolution due to heating from the decay of radiogenic nuclides, induces important $\mathrm{N_2}$ losses after the planetesimal formation. Measuring the $\mathrm{N_2}/ \mathrm{CO}$ ratio in comets is thus of great importance to decipher models of planetesimal formation and constrain the physical properties of the solar nebula at the time of their formation. From the detection of $\mathrm{N_2}$ by the ROSINA instrument in the coma of 67P, providing the only direct measurement of the $\mathrm{N_2}/ \mathrm{CO}$ in the coma of a comet so far, \cite{Rubin2015} confirmed that molecular nitrogen appears to be highly depleted in comets compared to the proto-solar value (a factor $\sim$ 25), suggesting that  cometary  grains formed  at low temperatures (below 50K). 

The detection of strong $\mathrm{N_2^+}$ emission lines in the coma of a relatively bright comet like R2 is a unique opportunity to add a measurement of the $\mathrm{N_2}/ \mathrm{CO}$ ratio for another comet, but also to try to determine the $\mathrm{^{14}N/ ^{15}N}$ isotopic ratio from a direct measurement in $\mathrm{N_2}$, the main reservoir of nitrogen in the early solar system, via $\mathrm{N_2^+}$. We thus decided to observe R2 with UVES, the high resolution optical spectrograph of the VLT, in order to secure a high quality spectrum of the comet over the full optical range. Hereafter, we present the results of those observations.


\section{Observations and data reduction}

\begin{table*}
\caption{Observing circumstances of comet C/2016 R2}
\label{observations}
\centering
\begin{tabular}{lccccccc}
\hline\hline
Date &  $\mathrm{r_h} (au)$ & $\mathrm{\dot{r}_h} (km s ^{-1})$ & $\Delta$ (au)& $\dot{\Delta}(km s ^{-1})$ & Exptime (s) & UVES Setup &  UVES Slit\\
\hline
2018-02-11T00:27:07.326 & 2.76 & -6.09 & 2.40 & 19.7 & 4800 & DIC1-390+580 & 0.44"$\times$8" - 0.44"$\times$12"\\
2018-02-13T00:46:23.196 & 2.76 & -5.97 & 2.43 & 19.9 & 4800 & DIC1-390+580 & 0.44"$\times$8" - 0.44"$\times$12" \\
2018-02-14T00:47:40.759 & 2.75 & -5.91 & 2.44 & 20.1 & 4800 & DIC1-390+580 & 0.44"$\times$8" - 0.44"$\times$12" \\
2018-02-15T00:23:16.493 & 2.75 & -5.85 & 2.45 & 20.1 & 3000 & DIC2-437+860 & 0.44"$\times$10" - 0.44"$\times$12" \\
2018-02-16T00:16:08.672 & 2.75 & -5.79 & 2.46 & 20.2 & 3000 & DIC2-437+860 & 0.44"$\times$10" - 0.44"$\times$12"\\

\hline
\end{tabular}
\end{table*}

We have been granted a total of 5 hours of Director's Discretionary Time, with the Ultraviolet-Visual Echelle Spectrograph (UVES) mounted on the ESO 8.2 m UT2 telescope to observe R2. We used two different UVES standard settings in order to cover completely the optical range of the spectrum: the dichroic \#1 (390+580) setting covering the range 326 to 454 nm in the blue and 476 to 684 nm in the red, and the dichroic \#2 (437+860) setting covering the range 373 to 499 nm in the blue and 660 to 1060 nm in the red. Because of clouds during the second observation performed with the dichroic \#1 setting, it was repeated, leading to a total of five exposures. The dates and observing circumstances of the 5 spectra are summarized in Table \ref{observations}. For both set-ups, we used a 0.44"-wide slit, providing a resolving power R~$\sim$~80~000. The smallest slit length of 8" samples about 14500 km at the distance of the comet ($\Delta=$2.5 au).

The data reduction was made using the ESO UVES pipeline, combined with custom routines to perform the extraction, cosmic rays removal, and correct for the Doppler shift due to the relative velocity of the comet with respect to the Earth. The spectra are calibrated in absolute flux using either the archived master response curve or the response curve determined from a standard star observed close to the science spectrum (both were used for R2 with no significant differences). More details can be found in the UVES ESO pipeline manual\footnote{\url{ftp://ftp.eso.org/pub/dfs/pipelines/uves/uves-pipeline-manual-22.17.pdf}}. Finally, the continuum, including the sunlight reflected by the cometary dust grains, was removed using a BASS2000 solar spectrum whose slope was corrected to match the one of the comet. As a consequence, the final comet spectrum we used contains the gas component only. More details regarding data reduction can be found in \cite{Manfroid2009}. 

In addition to the UVES spectra, we also used the TRAPPIST 60-cm telescopes \citep{Jehin2011} to monitor the general activity of the comet. TRAPPIST telescopes are equipped with sets of narrow-band filters allowing to isolate the emission lines of OH, NH, CN, $\mathrm{C_3}$, and $\mathrm{C_2}$ radicals, the $\mathrm{CO^+}$ ion, as well as several regions of the dust reflected continuum free from gas contamination \citep{Farnham2000}. More details about the TRAPPIST data reduction procedure can be found in \cite{Opitom2015}.

\section{Results}

The spectrum of R2 is very different from what we normally see in other comets at similar distances from the Sun. 
The optical spectrum of comets is usually dominated by the emission of radicals such as OH, NH, CN, $\mathrm{C_2}$, $\mathrm{C_3}$, or $\mathrm{NH_2}$. However, the spectrum of R2 is surprisingly dominated by several strong emission bands of the ions $\mathrm{CO^+}$ and $\mathrm{N_2^+}$, and to a lesser extent $\mathrm{CO_2^+}$.

\subsection{Narrow-band imaging}

\begin{figure*}
\centering
\includegraphics[width=7cm,height=4.5cm]{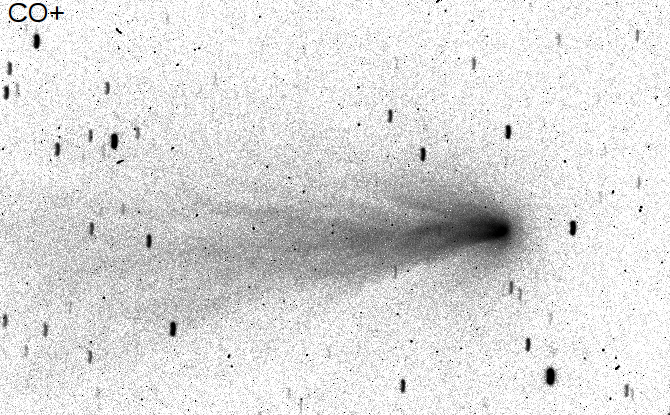} \includegraphics[width=7cm,height=4.5cm]{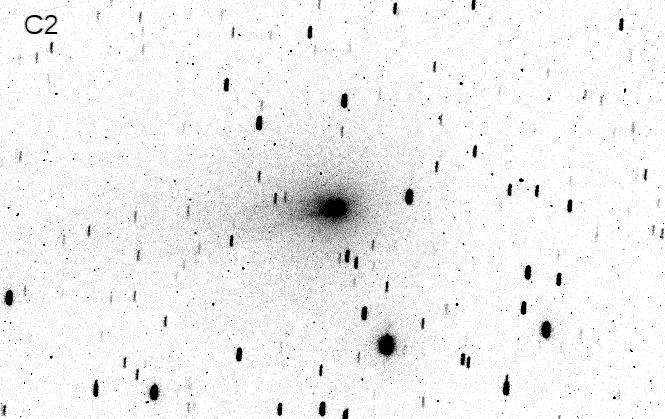}\\
\includegraphics[width=7cm,height=4.5cm]{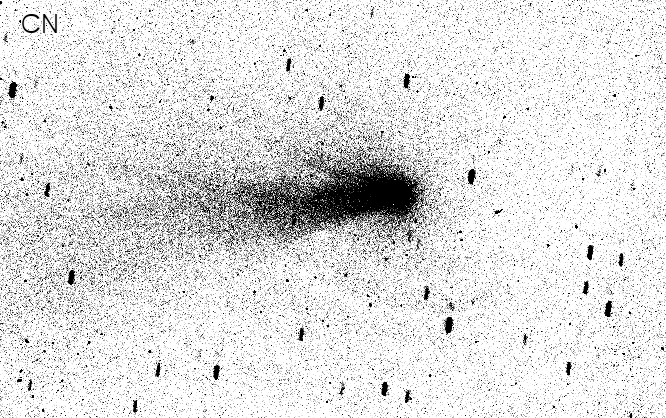} \includegraphics[width=7cm,height=4.5cm]{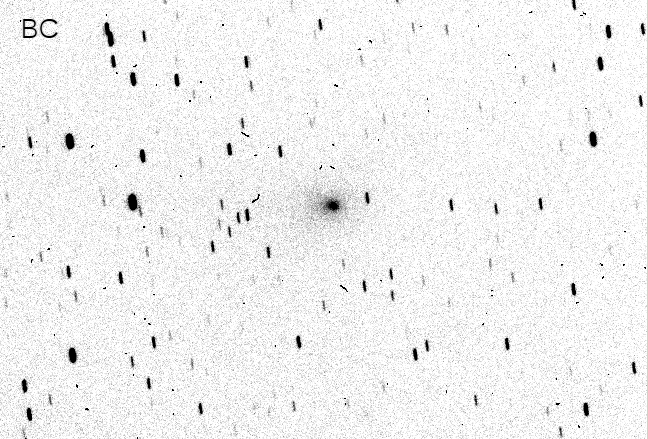}
\caption{TRAPPIST-South images of comet C/2016 R2 obtained with the $\mathrm{CO^+}$, $\mathrm{C_2}$, and CN narrow-band filters and TRAPPIST-North image obtained with the BC narrow-band filter on February 10 and 11, 2018 respectively. The images are oriented with North down and East left.}
\label{ImgCNCO+}
\end{figure*}

In Fig. \ref{ImgCNCO+}, we show images of R2 obtained in early February 2018 (at a heliocentric distance of 2.8 au) with TRAPPIST-South (TS) using the $\mathrm{CO^+}$, $\mathrm{C_2}$, and CN narrow-band filters, and an image obtained with TRAPPIST-North (TN) using the narrow-band BC filter. The $\mathrm{CO^+}$ coma shows a lot of structures changing night after night, and an enhancement in the anti-sunward direction, as expected for an ion. Instead of showing the usual large diffuse and symmetrical CN coma, the image through the CN filter has a morphology similar to the $\mathrm{CO^+}$ image, indicating that the CN filter is highly contaminated by an ion. This is due to the $\mathrm{N_2^+}$ emission band at 391.4 nm, which lies partially within the bandpass of the CN filter. The lack of the typical CN coma also indicates a low abundance of CN. Similarly, the $\mathrm{C_2}$ image shows structure, due to strong contamination by $\mathrm{CO^+}$ emission. Because of the contamination of the CN and $\mathrm{C_2}$ narrow-band filters by other emissions, and the low apparent abundance of those species, which will be discussed further later, we could not reliably compute CN and $\mathrm{C_2}$ production rates from narrow-band imaging. 

OH emission at 309 nm was not detected for any of the observing dates with TRAPPIST using the OH narrow band filter. Using simulated OH images produced from a Haser model \citep{Haser1957} and added to the real images, we could estimate the upper limits of the OH production rates for each observation. Images with the OH filter were taken with TN and TS from December 26, 2017 to February 11, 2018. They all agree with an upper limit of about $5\times10^{28}$ mol/s for the OH production rate. This is in agreement with (even though less constraining) the upper limit of $1.1\times10^{28}$ mol/s derived by \cite{Biver2018} from observations performed with the Nan\c{c}ay radio telescope between January 2 and March 26, 2018.

Using observations of the comet with narrow-band dust filters (BC and RC) performed with both TRAPPIST telescopes between December 26, 2017 and April 8, 2018 we derived $Af\rho$ values to constrain the dust content of the comet. Those values have been computed at a fixed nucleocentric distance of 10000km and are reported in Table \ref{Afrho}. From these observations, we compute a dust reddening between 15 and 24 \%/100nm.

\begin{table*}
\caption{$Af\rho$ values of comet C/2016 R2}
\label{Afrho}
\centering
\begin{tabular}{lccccc}
\hline\hline
Date &  $\mathrm{r_h} (au)$ & $\Delta$ (au) & $Af\rho$(BC) (cm) & $Af\rho$(RC) (cm)  & Telescope\\
\hline
2017-12-26 & 2.96 & 2.06 & $752\pm4$ & $1248\pm8$ & TN\\
2017-12-27 & 2.95 & 2.06 & - & $1158\pm10$ & TN\\
2017-12-27 & 2.95 & 2.06 & $622\pm11$ & $944\pm6$ & TS\\
2018-01-04 & 2.92 & 2.08 & $631\pm5$ & $1051\pm9$ & TN\\
2018-01-11 & 2.88 & 2.12 & $536\pm4$ & $1010\pm10$ & TN\\
2018-01-20 & 2.84 & 2.19 & $520\pm4$ & $888\pm3$ & TN\\
2018-02-11 & 2.76 & 2.41 & $503\pm6$ & $829\pm5$ & TN\\
2018-02-13 & 2.75 & 2.43 & $458\pm6$ & $882\pm6$ & TN\\
2018-03-19 & 2.65 & 2.83 & $408\pm7$ & $802\pm8$ & TN\\

\hline
\end{tabular}
\end{table*}

\subsection{Detected and non-detected species}

\begin{figure*}
\centering
\includegraphics[width=14cm]{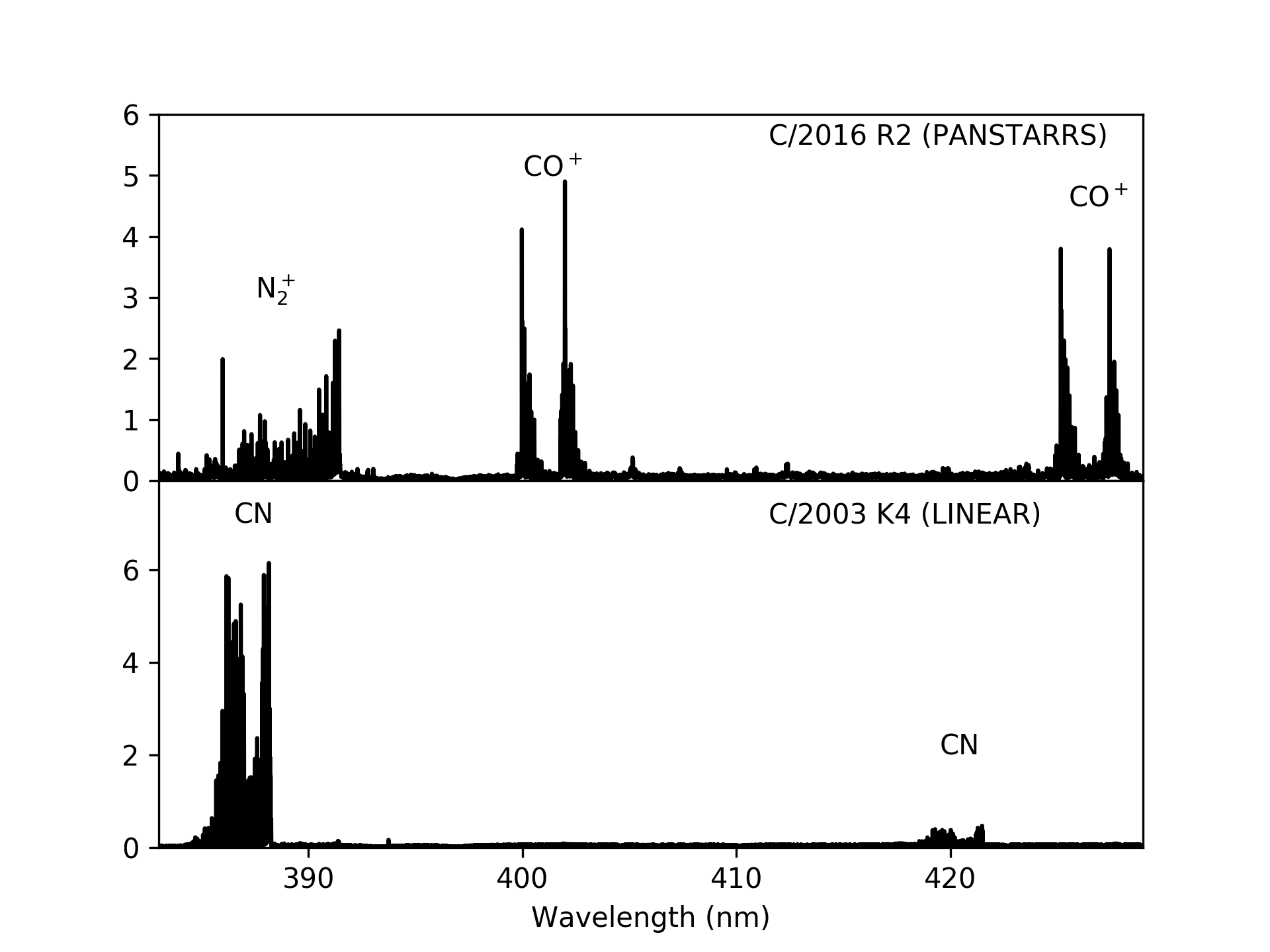}
\caption{Comparison of the UVES spectra of C/2016 R2 at 2.8 au and C/2003 K4 at 2.6 au over the 383.0 to 429.0 nm wavelength range. The y-axis has arbitrary units and has been chosen independently for both spectra so bright emission features appear with similar intensities.}
\label{SpectR2K4}
\end{figure*}

\begin{figure*}
\centering
\includegraphics[width=14cm]{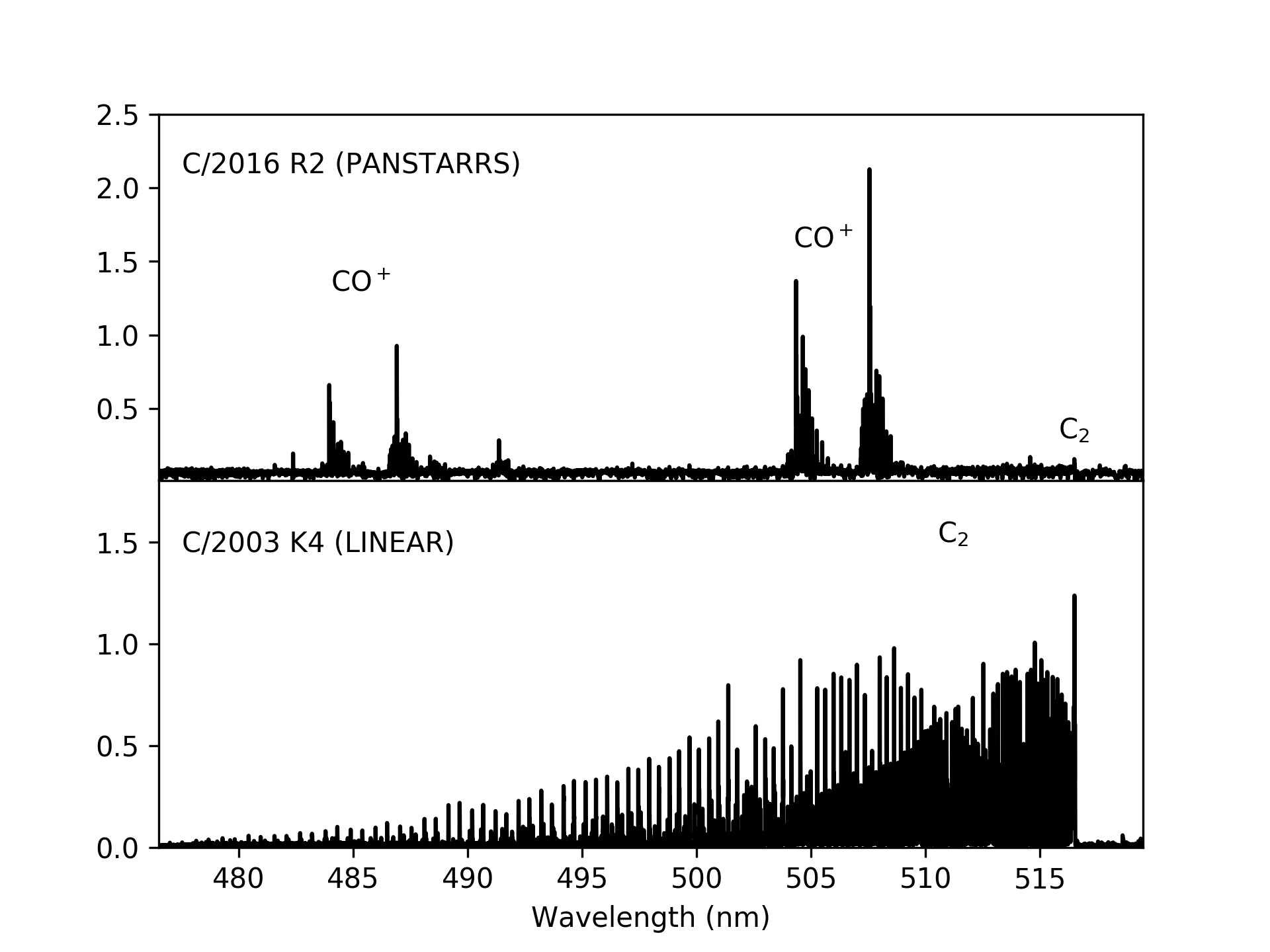}
\caption{Comparison of the UVES spectra of C/2016 R2 at 2.8 au and C/2003 K4 at 2.6 au over the 476.5 to 519.5 nm wavelength range. The y-axis has arbitrary units and has been chosen independently for both spectra so bright emission features appear with similar intensities.}
\label{SpectR2K4C2}
\end{figure*}

Compared to previous observations of this comet with other high resolution spectrographs at optical wavelengths, UVES spectra have the advantage of covering the whole optical range while benefiting of being mounted on an 8m-class telescope, allowing to detect fainter emission bands or species. With this spectrum, we confirm the peculiarity of R2, as illustrated in Fig. \ref{SpectR2K4} and \ref{SpectR2K4C2}, where it is compared to the spectrum of comet C/2003 K4 (LINEAR) obtained with the same instrument and spectral resolution and at comparable heliocentric and  geocentric distances \citep{Manfroid2005}. The spectrum of C/2003 K4 is dominated by strong CN emission lines around 390 nm and $\mathrm{C_2}$ between 450 and 520 nm, while the spectrum of R2 is dominated by $\mathrm{N_2^+}$ emission below 392 nm and by $\mathrm{CO^+}$ in the 400-620 nm region as already shown by \cite{Cochran2018}.  This is also consistent with the CN narrow-band images of R2 obtained with TRAPPIST being contaminated by $\mathrm{N_2^+}$ emission lines.

\begin{figure}
\centering
\includegraphics[width=9.42cm]{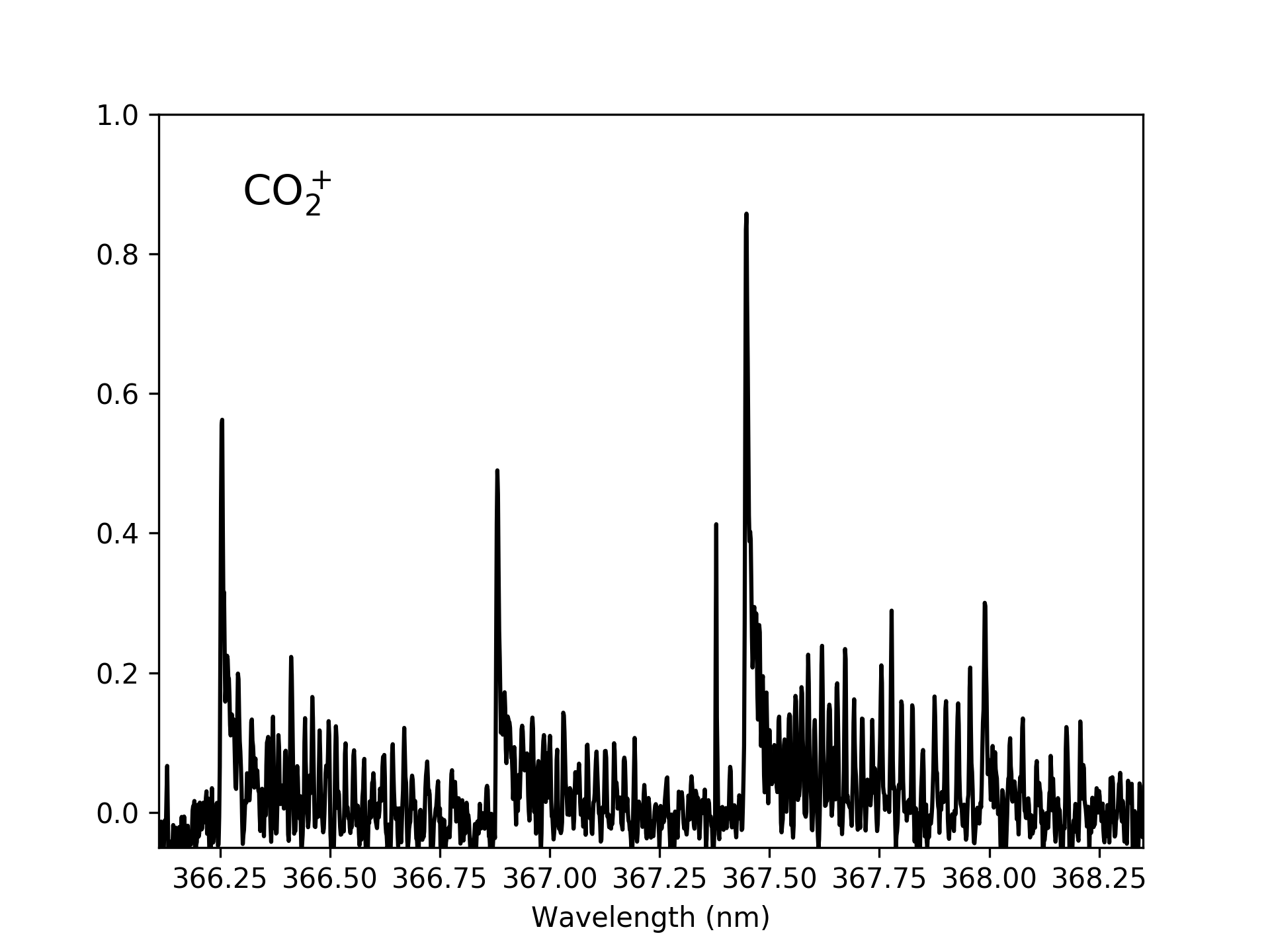}
\caption{Spectrum of C/2016 R2, showing several $\mathrm{CO_2^+}$ bands: (0,0,0) $\mathrm{\tilde{A}^2\Pi_{u,3/2}-(1,0,0)\tilde{X}^2\Pi_{g,3/2}}$ at 366.3 and 366.9nm and (0,0,0) $\mathrm{\tilde{A}^2\Pi_{u,1/2}-(1,0,0)\tilde{X}^2\Pi_{g,1/2}}$ at 367.4nm. The y-axis has arbitrary units.}
\label{SpectCO2+}
\end{figure}

\begin{figure}
\centering
\includegraphics[width=9.42cm]{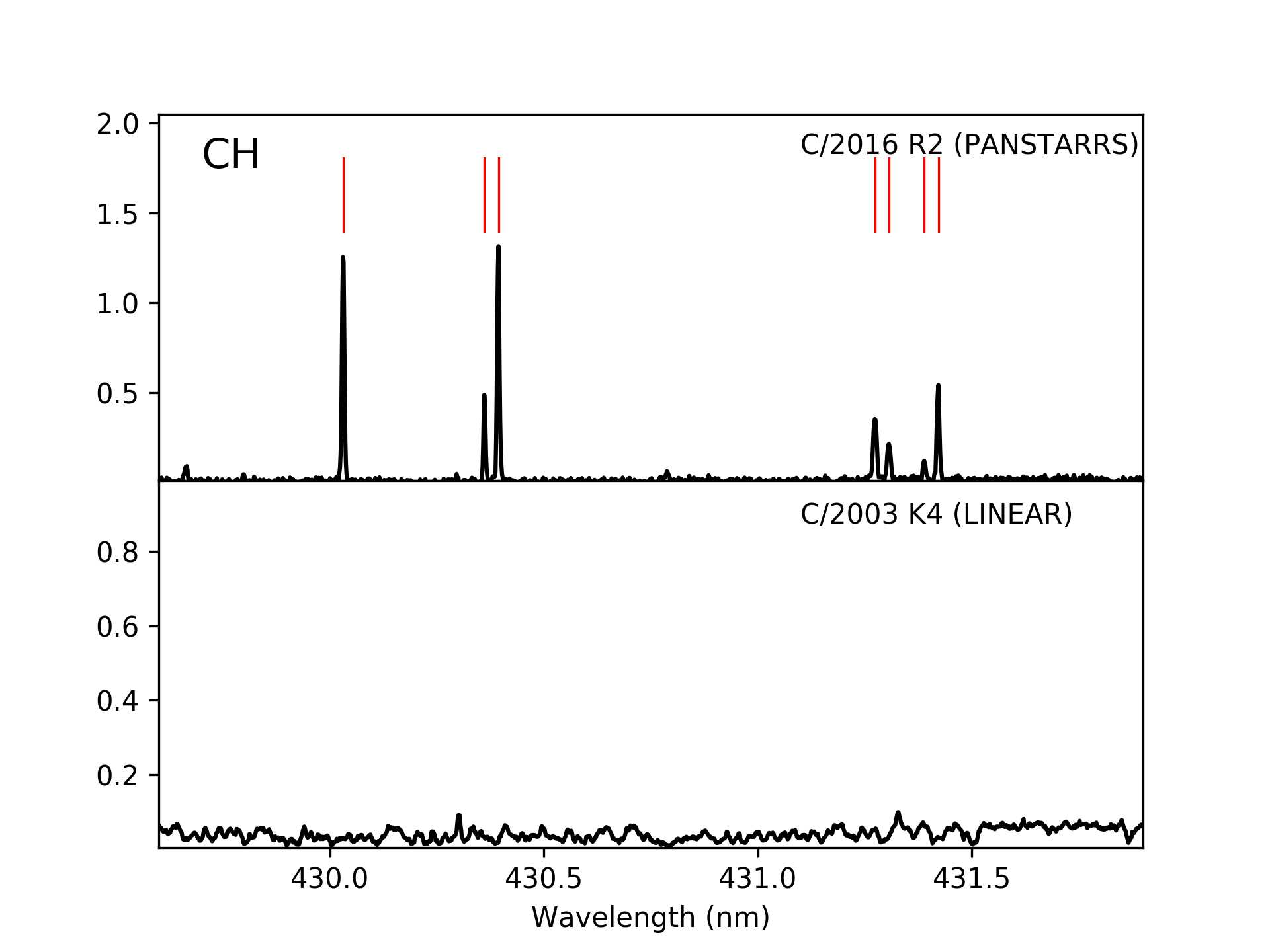}
\caption{Comparison of the UVES spectra of C/2016 R2 at 2.8 au and C/2003 K4 at 2.6 au over the 429.5 to 432 nm wavelength range, showing the CH lines from the (0,0) band in R2 (indicated by the vertical red lines) and their absence in K4. The y-axis has arbitrary units and has been chosen to display both spectra on a similar scale.}
\label{PlotCH}
\end{figure}

The list of species identified in the UVES spectra is given in Table~\ref{Detect}, along with their corresponding transitions. The strongest signatures are those of $\mathrm{N_2^+}$, $\mathrm{CO^+}$, and $\mathrm{CO_2^+}$. In addition to the (4,0), (3,0), (2,0), (4,2), (3,2), (1,0), (2,1), and (1,1) $\mathrm{A^2\Pi-X^2\Sigma}$ $\mathrm{CO^+}$ bands reported by \cite{Cochran2018}, we detect the (6,0), (7,1), (5,0), (0,0), (0,1), (1,2), and (0,2) bands of the same system. The (0,0), (1,1), and (0,1) bands of the $\mathrm{N_2^+}$ first negative group ($\mathrm{B^2\Sigma_u^+-X^2\Sigma_g^+}$) are also detected, and we have evidence for the detection of the (0,2) band. 
Finally, besides $\mathrm{N_2^+}$ and $\mathrm{CO^+}$, we also detect several $\mathrm{CO_2^+}$ bands, as illustrated in Fig. \ref{SpectCO2+}. In addition to those three ions, which are dominating the spectrum of R2, we detect some of the usual CN, $\mathrm{C_3}$, and $\mathrm{C_2}$ emission lines, but those are much fainter than in other comets at the same distance (see Fig. \ref{PlotQCN} for the CN (0,0) band and the very faint $\mathrm{C_2}$ bandhead visible in Fig. \ref{SpectR2K4C2}). In Fig. \ref{PlotCH}, the CH (0,0) band is clearly detected, while it was not in the coma of comet C/2003 K4 (LINEAR) observed at a similar heliocentric distance, indicating that R2 seems to be enriched in CH. We also have a tentative identification of a few $\mathrm{CH^+}$ lines.

As interesting are species that are usually seen in the coma of similarly bright comets but that could not be detected spectra of R2. We searched for OH emission lines, to get an indication of the water content of the comet. We looked for matching lines from the $\mathrm{A^2\Sigma^+-X^2\Pi_i}$ (0,1) band around 345 nm, without success. Unfortunately, the setup used did not cover the strongest (0,0) OH emission band around 309 nm. Similarly, we also searched for $\mathrm{OH^+}$ lines around 356.5 nm without success.

\begin{figure}
\centering
\includegraphics[width=9.4cm]{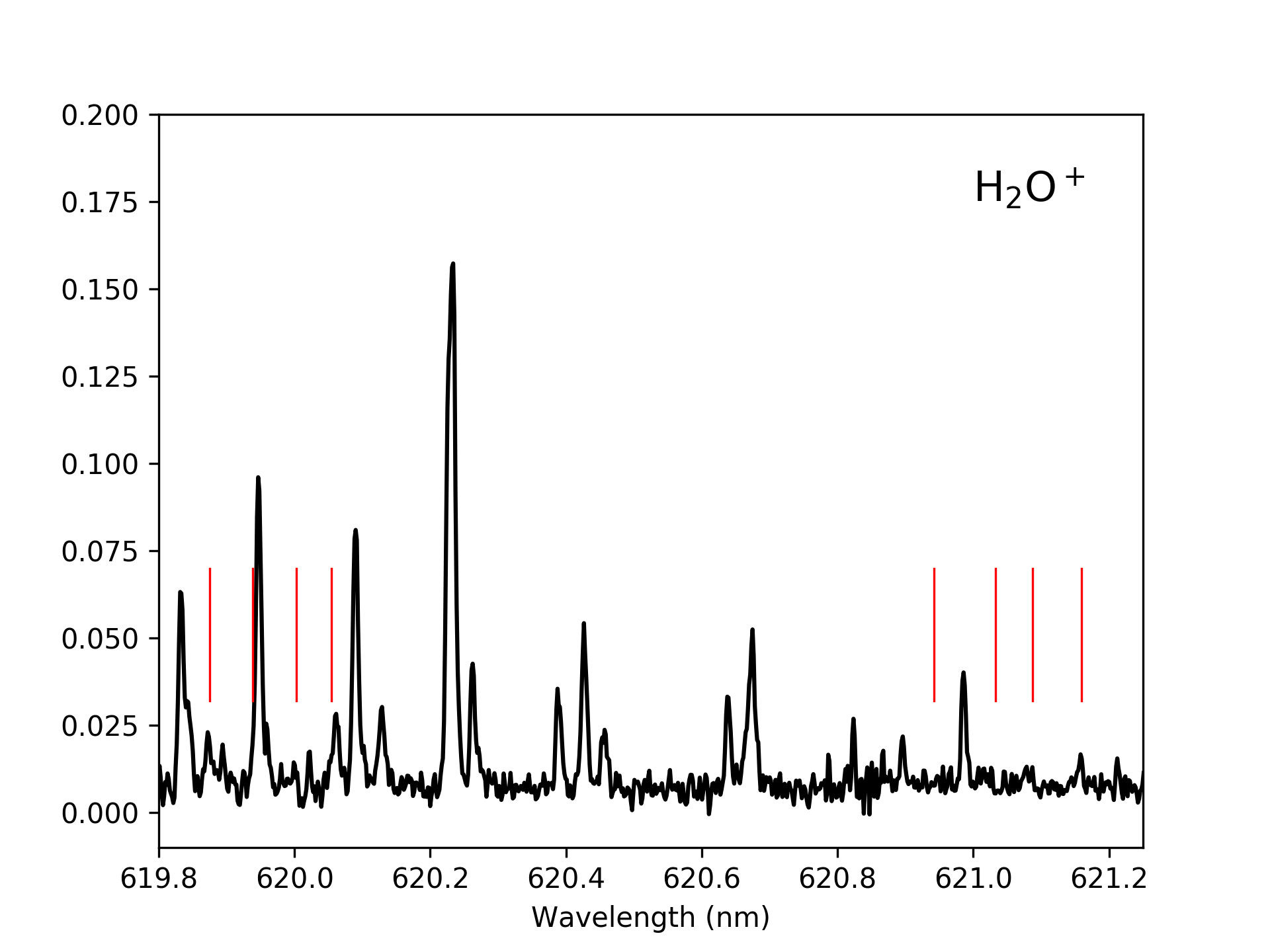}
\caption{Spectrum of C/2016 R2 showing the region of $\mathrm{H_2O^+}$ (8-0) band. The red tick marks indicate the wavelength of several of the brightest $\mathrm{H_2O^+}$ lines, which are not clearly detected in C/2016 R2. Most lines detected in this wavelength range belong to the $\mathrm{CO^+}$ (0-2) band. The y-axis has arbitrary units.}
\label{PlotH20}
\end{figure}

\begin{figure}
\centering
\includegraphics[width=9.4cm]{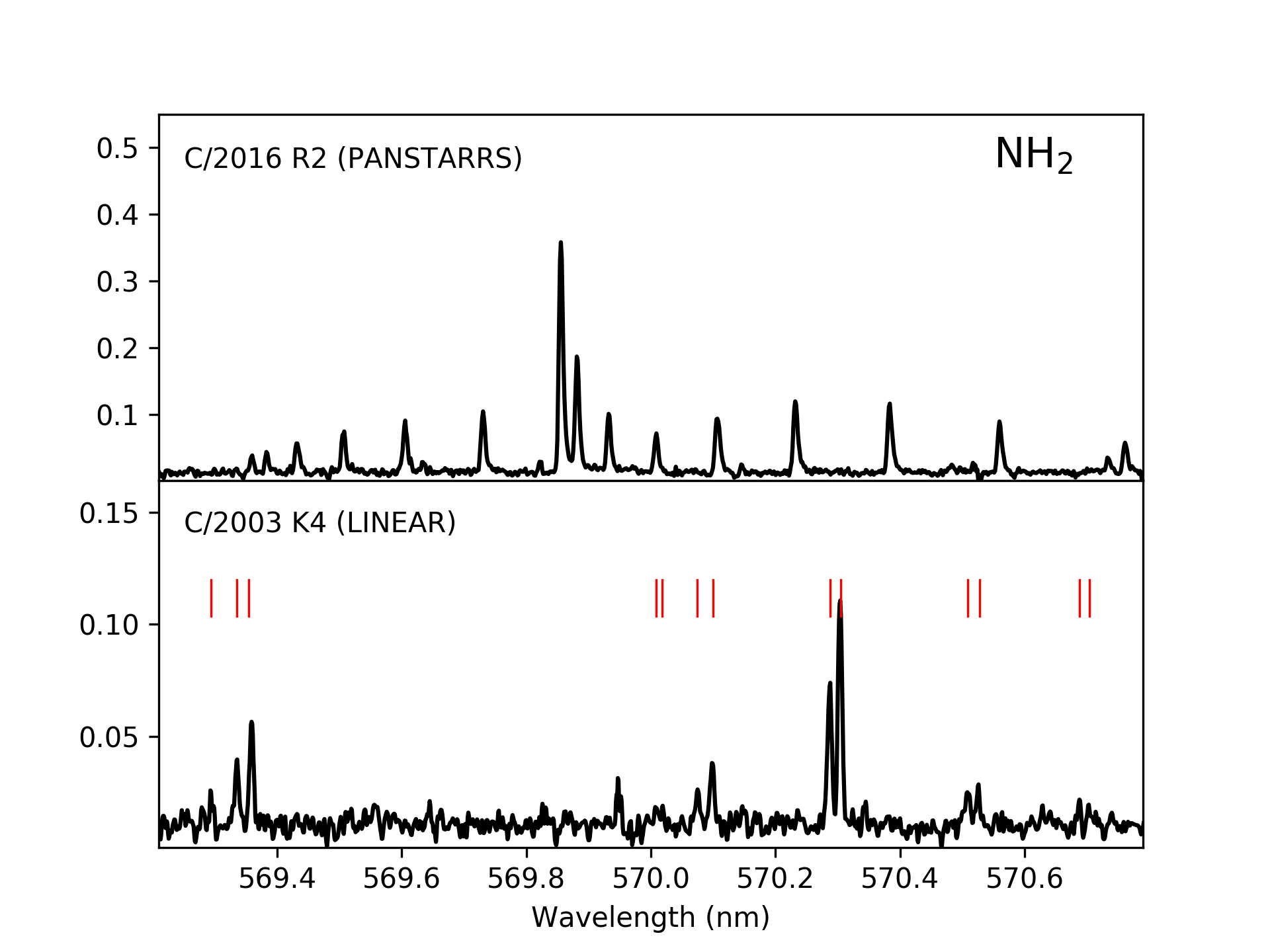}
\caption{Comparison of the UVES spectra of comet C/2016 R2 at 2.8 au and C/2003 K4 at 2.6 au over the 569.3 to 570.7 nm wavelength range, showing the detection of $\mathrm{NH_2}$ lines in K4 (indicated by the vertical red lines, those lines belong to the A(0,10,0) - X(0,0,0) band) and their non-detection in R2. The y-axis has arbitrary units and has been chosen to display both spectra on a similar scale.}
\label{PlotNH2}
\end{figure}

$\mathrm{H_2O^+}$ has emission lines in the 580-755 nm region, which are regularly detected in the coma of comets (see for example \cite{Cochran2002}). Even though, as illustrated in Fig. \ref{PlotH20}, two faint lines around 618.42 and 619.88 nm (and maybe a third one around 614.68 nm) could potentially match the wavelength of $\mathrm{H_2O^+}$ emission lines, there is no clear evidence for the detection of $\mathrm{H_2O^+}$ in the coma of R2. Ultimately, we looked for $\mathrm{NH_2}$ emission lines around 570 nm, which we would have expected in such a bright comet at that distance from the Sun. Those lines were for example easily detected in C/2003 K4 (LINEAR) observed under similar circumstances. In Fig. \ref{PlotNH2}, we show several $\mathrm{NH_2}$ lines detected in K4, which are not detected in the coma of R2. Most lines visible in Fig. \ref{PlotNH2} in the spectrum of R2 belong to the (1,2) $\mathrm{CO^+}$ band.

Other observations of R2 at optical or radio wavelengths have been performed. \cite{deValBorro2018} measured a CO production rate between 1.09 and $1.44\times 10^{29}$ mol/s from observations performed with the Arizona Radio Observatory's Submillimeter Telescope during the period January 10-16, 2018, consistent with what was measured by \cite{Biver2018} from observations performed with IRAM on January 23-24 2018. \cite{deValBorro2018} did not detect $\mathrm{CH_3OH}$, HCN, CS, $\mathrm{H_2CO}$, or $\mathrm{HCO^+}$. \cite{Wierzchos2018} also reported that the comet was depleted in HCN compared to CO. This is consistent with the faintness of the CN (0-0) band in the UVES spectrum. Recently, \cite{Biver2018} reported the detection of small amounts of HCN as well as the non-detection of OH at radio wavelengths, and confirmed that the comet is strongly depleted in $\mathrm{H_2O}$, $\mathrm{CH_3OH}$, $\mathrm{H_2CO}$, HCN, amd $\mathrm{H_2S}$ with respect to CO compared to comets observed at similar heliocentric distances.

From observations performed with the 2.7 m telescope of the McDonald Observatory in December 2017 at optical wavelengths, \cite{Cochran2018} could not detect CN, $\mathrm{C_2}$, $\mathrm{C_3}$, or CH, while we see those species about two months later. Although this might be due to the fact that when their observations were performed, the comet was further away from the Sun and less active, this is most probably because UVES is mounted on an 8-m telescope which provides a much higher sensitivity.

\cite{Kumar2018} reported the detection of $\mathrm{H_2O^+}$ in the coma of R2 using the LISA spectrograph mounted on the 1.2~m telescope of the MIRO observatory, from observations performed on January 25, 2018, while we do not convincingly detect $\mathrm{H_2O^+}$ emission lines two weeks later. However, the spectrum presented by \cite{Kumar2018} was obtained at much lower spectral resolution. As illustrated in Fig. \ref{PlotH20}, in the wavelength range of the $\mathrm{H_2O^+}$ emission lines, we detect numerous emission lines, either belonging to $\mathrm{CO^+}$, identified as sky lines, or unidentified. At low resolution, those could be mistaken for $\mathrm{H_2O^+}$, so we believe that the detection of $\mathrm{H_2O^+}$ by \cite{Kumar2018} should be taken with caution. We also note that $\mathrm{H_2O^+}$ is not detected by \cite{Cochran2018} in their observations performed in November and December 2017.

\begin{table*}
\caption{Detected species in the coma of comet C/2016 R2}
\centering
\begin{tabular}{|l|p{11cm}|p{4cm}|}
\hline\hline
Species &  Detected bands & Respective band/line wavelengths \\
\hline
$\mathrm{CO^+}$ & (6,0), (7,1), (5,0), (4,0), (3,0), (2,0), (4,2), (1,0), (2,1), (3,2), (0,0), (1,1), (0,1), (1,2), (0,2) $\mathrm{A^2\Pi-X^2\Sigma}$ bands & 341, 351, 359, 378, 400, 425, 452,  454, 468, 484, 490, 504, 547, 567, and 619~nm \\

$\mathrm{N_2^+}$ & (0,0), (1,1), (0,1) $\mathrm{B^2\Sigma_u^+-X^2\Sigma_g^+}$ bands and a few matching lines for the (0,2) band & 391, 389, 428, and 471 nm \\ 

$\mathrm{CO_2^+}$ & (0,0,0))$\mathrm{\tilde{A}^2\Pi_{u,3/2}-(0,0,0)\tilde{X}^2\Pi_{g,1/2}}$, (0,0,0)$\mathrm{\tilde{A}^2\Pi_{u,3/2}-(0,0,0)\tilde{X}^2\Pi_{g,3/2}}$ &   \\
 & the (0,0,0) $\mathrm{\tilde{A}^2\Pi_{u,3/2}-(1,0,0)\tilde{X}^2\Pi_{g,3/2}}$, and the (0,0,0) $\mathrm{\tilde{A}^2\Pi_{u,1/2}-(1,0,0)\tilde{X}^2\Pi_{g,1/2}}$ bands & 351, 350, 366, and 366 nm \\

$\mathrm{CH^+}$ & Tentative detection of the (0,0) $\mathrm{A^1\Pi-X^1\Sigma^+}$ band & 423 nm \\
 
 $\mathrm{CH}$ & (0,0) $\mathrm{A^2\Delta-X^2\Pi}$ band and tentative detection of the (0,0) $\mathrm{B^2\Sigma^--X^2\Pi}$ band & 431 and 389 nm  \\
 
CN           & (0,0),(1,0) $\mathrm{B^2\Sigma^+-X^2\Sigma^+}$ bands and tentative detection of the (1,1) band & 388, 422, and 388nm \\

$\mathrm{C_2}$ & (0,0) $\mathrm{d^3\Pi_g-A^3\Pi_u}$ band bandhead (Swan System) & 517 nm \\

$\mathrm{C_3}$ & (0,0,0)-(0,0,0) and (0,2,0)-(0,0,0) bands and tentative detection of the (0,0,0)-(0,2,0)  $\mathrm{\tilde{A}^1\Pi_u-\tilde{X}^1\Sigma_g^+}$ band & 405, 392, and 407 nm  \\ 

$\mathrm{[OI]}$ & Three forbidden oxygen lines & 557.7339, 630.0304, and 636.3776 nm\\

$\mathrm{[NI]}$ & Two forbidden nitrogen lines & 519.7900 and 520.0256 nm \\ 

$\mathrm{[CI]}$ & Three forbidden carbon lines & 872.712, 982.411, and 985.024 nm \\

\hline
\end{tabular}
\tablefoot{OH, $\mathrm{OH^+}$, $\mathrm{H_2O^+}$, and $\mathrm{NH_2}$ are not detected.}\\

\label{Detect}

\end{table*}

\subsection{The dust-to-gas ratio}

\begin{figure}
\centering
\includegraphics[width=9.42cm]{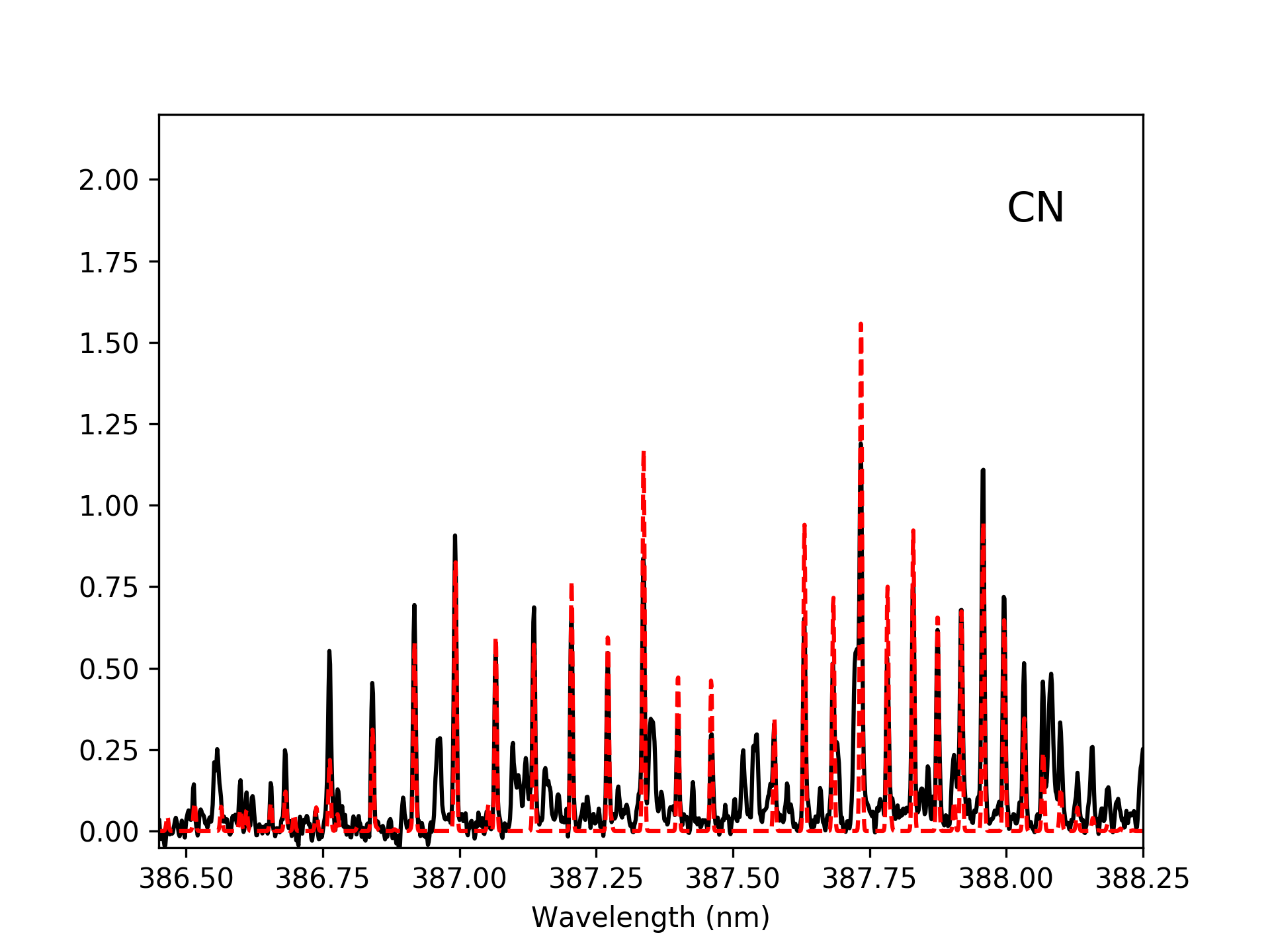}
\caption{Spectrum of C/2016 R2 showing the CN (0,0) band in black, overlapped with our best CN model (red dashed line). The y-axis has arbitrary units.}
\label{PlotQCN}
\end{figure}

Determining absolute gas production rates from high resolution spectra using a small slit is not straightforward, as a correction for slit losses has to be applied. The CN production rate is particularly difficult to measure in the case of R2, because of the $\mathrm{N_2^+}$ (0,0) band partly overlapping with the CN (0,0) band. However, using a model of the CN (0,0) band, combined with a Haser profile to determine the part of the flux encompassed in the slit (and scalelengths from \cite{AHearn1995}), we estimate a CN production rate of $(3\pm1)\times10^{24}$ mol/s. To perform this measurement, we use the first spectrum obtained on February 11 under clear sky conditions. The CN (0,0) band, along with our best model are displayed on Fig. \ref{PlotQCN}.

Using this estimate of the CN production rate and the $Af\rho$ measured from TN observations in the BC filter made on the same date, we can compute the $Af\rho$/Q(CN) ratio, as a proxy of the dust-to-gas ratio of the comet. We obtain a $log[Af\rho$/Q(CN)]=$-21.78\pm0.14$. Comparing this value to the comets presented in the data set of \cite{AHearn1995}, and also to more than 30 comets observed with TRAPPIST over the past 6 years, this value is actually higher than for most comets, indicating either a dust-rich, or CN-poor comet. We already mentionned previously, and it is illustrated in Fig. \ref{SpectR2K4}, that the CN emission in the coma of R2 is faint compared to other comets observed at similar heliocentric distances. Comparing the measured $Af\rho$ value to comets observed at similar heliocentric distance by \cite{AHearn1995}, R2 is within the range of measured $Af\rho$ values. \cite{Biver2018} compared the $Af\rho$/CO ratio of R2 to comets at similar distances and with available CO abundances and found it particularly low. The question is then: is R2 rich in CO, and/or poor in dust? Between 2.5 and 3 au from the Sun, $\mathrm{H_2O}$, CO, and $\mathrm{CO_2}$ can all significantly contribute to the comet activity, and their relative contributions vary from comet to comet (see for example \cite{Ootsubo2012}). A thorough investigation of the dust-to-gas ratio of comets observed at similar distances, accounting for the contribution of the 3 species would allow to decipher whether R2 is actually dust-poor, or just very rich in CO. However, simultaneous (or close to simultaneous) and reliable measurements of the $\mathrm{H_2O}$, CO, and $\mathrm{CO_2}$ abundances in comets observed between 2.5 and 3 au are rare, making such a comparison difficult.

\subsection{[NI], [OI] and [CI] forbidden atomic lines}

\begin{figure}
\centering
\includegraphics[width=9.5cm]{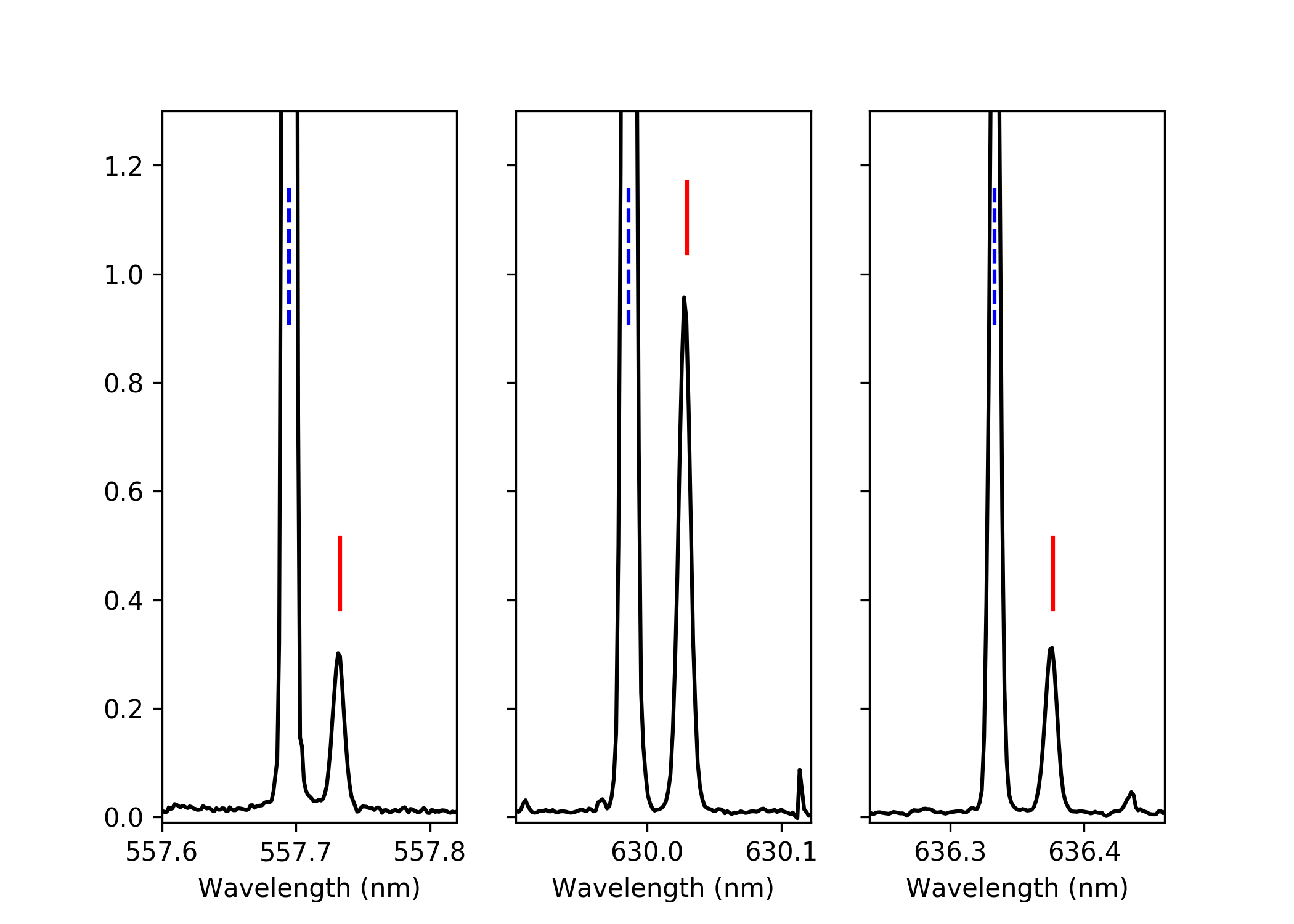}
\caption{The 3 forbidden oxygen [OI] lines in comet C/2016 R2 at 557.73 nm (left) for the green line, and 630.03 nm (middle) and 636.38 nm (right) for the red doublet. The cometary oxygen lines are identified by red tick marks while the dashed blue tick marks indicate the corresponding telluric lines. The y-axis is in arbitrary units.}
\label{PlotOI}
\end{figure}

Aside from the molecules mentioned above, we observe emissions from atomic species. The green forbidden oxygen line, at 557.73 nm and the red doublet at 630.03 and 636.38 nm are clearly identified (see Fig. \ref{PlotOI}), shifted from the corresponding telluric lines due to the geocentric velocity of the comet. Atomic oxygen in the coma of comets is mainly produced by the photo-dissociation of $\mathrm{H_2O}$, $\mathrm{CO}$, and $\mathrm{CO_2}$, and, as recently discovered, $\mathrm{O_2}$ \citep{Bieler2015}. It has been shown that mixing ratios of $\mathrm{CO_2/H_2O}$ and $\mathrm{CO/H_2O}$, which are usually very difficult to determine from the ground, can be derived from observations of the ratio of the green oxygen line to the red doublet (G/R ratio) \citep{Festou1981,McKay2013,Decock2013,Decock2015}.

For R2, we measure a G/R ratio of 0.23 on the average spectrum summing all the flux along the  slit. \cite{Decock2013} reported a mean value of 0.11 for their sample of 11 comets, which is about twice lower than what we measure for R2. However, both \cite{Decock2013} and \cite{McKay2015} reported that comets observed at larger heliocentric distance have higher values of the G/R ratio. For example, \cite{Decock2013} reported a ratio of 0.3 for comet C/2001 Q4 (NEAT) observed at 3.73 au and ratios of 0.14 and 0.20 for comet C/2009 P1 (Garradd) observed at 2.9 au and 3.25 au respectively. \cite{McKay2015} measured a ratio of 0.16 for comet C/2009 P1 (Garradd) observed at 2.88 au and \cite{McKay2012} reported a value of 0.24 for comet C/2006 W3 (Christensen) observed at 3.13 au. The G/R ratio measured for R2 is thus consistent with the ratio measured for other comets observed at large heliocentric distances. The increase of the G/R ratio with the heliocentric distance has been interpreted by the fact that the water sublimation becomes less important at distances larger than 2.5 au so that other volatiles, such as CO and $\mathrm{CO_2}$ start to contribute more significantly to the comet activity. 
According to \cite{Decock2013}, while for comets observed below 2 au the observed mean G/R ratio of 0.09 is consistent with $\mathrm{H_2O}$ being the main molecule photo-dissociated to produce oxygen atom in metastable state, the high G/R ratio measured at large heliocentric distance indicates that at those distances $\mathrm{CO}$ and $\mathrm{CO_2}$, are contributing. R2 was observed at $\sim$2.8 au from the Sun and its spectrum shows strong emission lines of $\mathrm{CO^+}$ and $\mathrm{CO_2^+}$, while no OH or $\mathrm{H_2O^+}$ emission lines are convincingly detected. This is consistent with CO and $\mathrm{CO_2}$ contributing to drive the activity of R2, implying a high G/R ratio. 

\begin{figure}
\centering
\includegraphics[width=9cm]{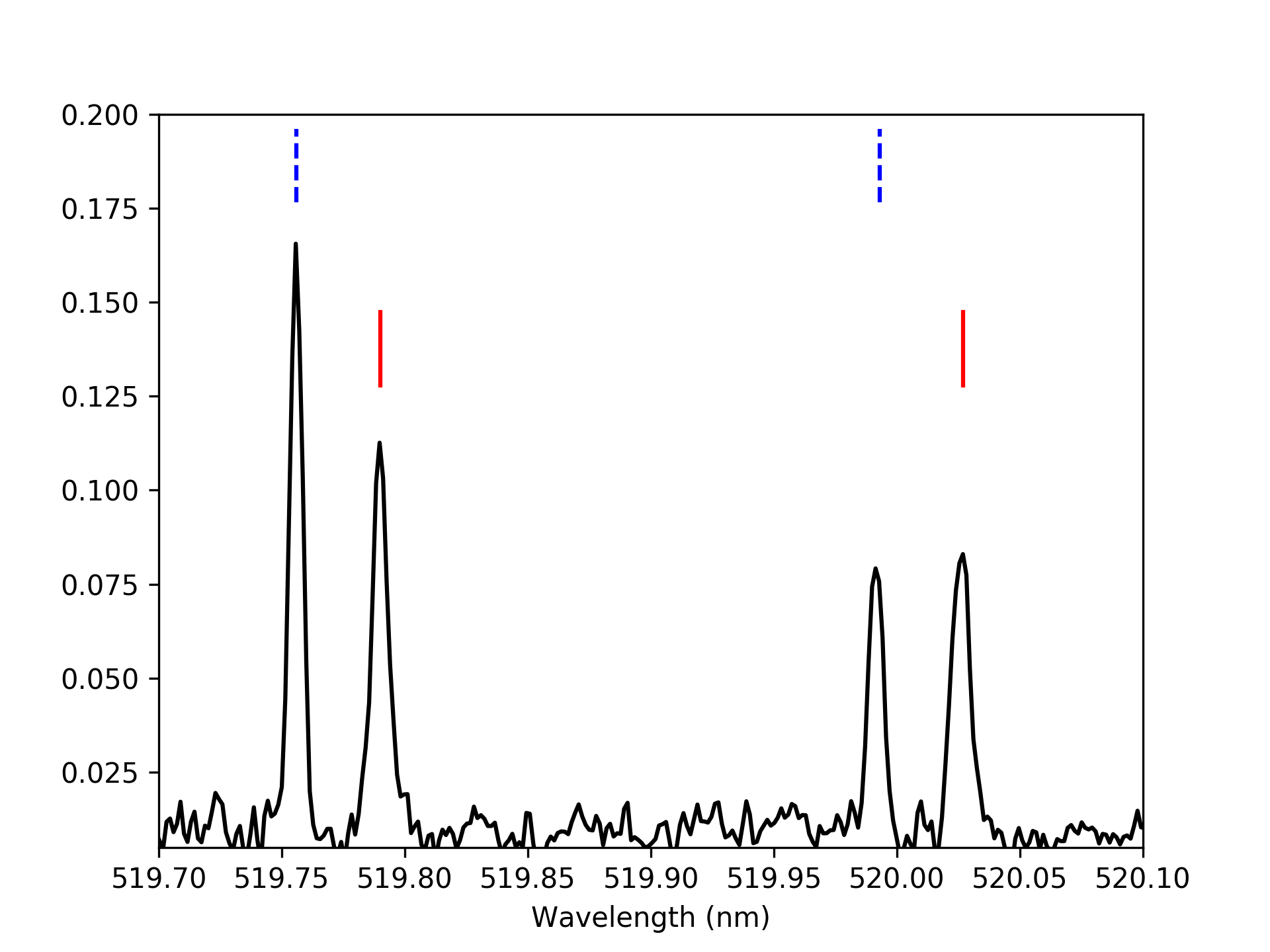}
\caption{Spectrum of C/2016 R2 between 519.7 and 520.1 nm. The full red tick marks indicate the two cometary nitrogen forbidden lines at 519.79 and 520.02 nm, while the dashed blue tick marks indicate the corresponding telluric lines. The y-axis is in arbitrary units.}
\label{PlotN}
\end{figure}

More surprisingly, we detect two lines at 519.79 and 520.02 nm, that we identify as being forbidden nitrogen lines \citep{Wiese1966}, see Fig. \ref{PlotN}. To our knowledge, this is the first time those lines are detected in the coma of a comet, even if \cite{Singh1991} mentioned that an unidentified line at 520.1 nm detected in several comets could match the forbidden nitrogen transitions. Those lines are produced by the decay of nitrogen in a metastable state $\mathrm{N(^2D^0)}$. 
Nitrogen in this metastable state could be produced in the coma of comets by photodissociation of CN for example \citep{Singh1991}. In the Earth atmosphere, nitrogen in the $\mathrm{N(^2D^0)}$ state is produced by electron impact dissociation of $\mathrm{N_2}$, electron impact dissociative ionization of $\mathrm{N_2}$, or dissociative electron recombination of $\mathrm{N_2^+}$ among others \citep{Rees1985}. \cite{Cravens1978} have studied airglow phenomenon in the inner coma of comets and, even though this is not included in their model, they suggest that forbidden nitrogen emission at 520 nm could result from the dissociative excitation of $\mathrm{N_2}$. Given the high $\mathrm{N_2^+}$ and the low CN abundances in R2, the dissociative electron recombination of $\mathrm{N_2^+}$, electron impact dissociative ionization of $\mathrm{N_2}$, and/or the electron impact dissociation of $\mathrm{N_2}$ are probably the main contributors to the production of nitrogen atoms in this metastable state. In Fig. \ref{PlotN}, we can see that the ratio between the [NI] 519.8 nm and 520.0 nm lines is different for the telluric lines ($\sim$2) and the cometary lines ($\sim$1.2). However, this ratio is expected to vary with the electron density \citep{Dopita1976} and observations of several planetary nebulae by \cite{Aller1970} for example show variations between 0.84 and 1.95. Measurements of the ratio in the Earth atmosphere have been performed by \cite{Sharpee2005}, which found a mean value of 1.759, close to what we measure for telluric lines.

\begin{figure}
\centering
\includegraphics[width=9.5cm]{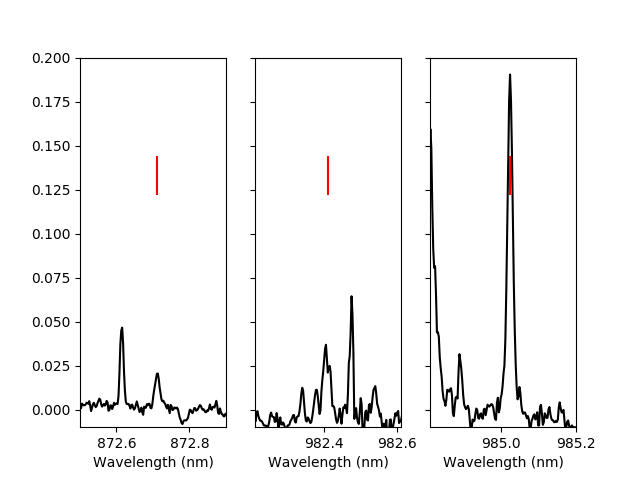}
\caption{Spectrum of C/2016 R2 corrected from telluric absorption, and covering the region of the [CI] forbidden lines at 872.7 nm (left), 982.4 nm (middle), and 985.0 nm (right). The red tick marks indicate the position of the carbon forbidden lines. The y-axis is in arbitrary units.}
\label{PlotCI}
\end{figure}

Forbidden [CI] lines at 872.7, 982.4 and 985.0 nm are emitted from $\mathrm{C(^1D)}$, produced by the photo-dissociation of neutral C-bearing species, electron-impact on the carbon ground-state $\mathrm{C(^3P)}$, or dissociative recombination of C-bearing ions such as $\mathrm{CO^+}$. Even though carbon forbidden lines are regularly detected in the coma of comets at UV wavelengths, the 985.0 nm line has been claimed to be detected previously only once, in the coma of comet Hale-Bopp by \cite{Oliversen2002} using a Fabry-Perot technique. In the spectrum of R2, we first observed a relatively strong line at 985.0 nm and a fainter one at 872.7 nm, as illustrated in Fig. \ref{PlotCI} (right and left, respectively), but a priori no emission at 982.4 nm. According to \cite{Hibbert1993}, the intensity ratio between the 985.0 and 982.4 nm lines should be about 3, so that we should easily detect the 982.4 nm line. After correction for a strong underlying telluric absorption, using the Molecfit tool \citep{Smette2015}, the 982.4 nm forbidden carbon line is clearly detected (Fig. \ref{PlotCI}, middle part), thought its position, width, and intensity cannot be accurately determined.

A more thorough modeling of all forbidden lines and their interpretation in terms of mixing ratio of the main contributers to comet activity is out of the scope of this work and will be done in a subsequent paper.

\subsection{The $\mathrm{N_{2}^+/CO^+/CO_2^+}$ abundance ratios} 

The $\mathrm{N_{2}^+}$, $\mathrm{CO^+}$, and $\mathrm{CO_2^+}$ ions have strong detected emissions, such that we can compute the ratios between those species. For $\mathrm{CO_2^+}$, we use the (0,0,0)$\mathrm{\tilde{A}^2\Pi_{u,3/2}-(0,0,0)\tilde{X}^2\Pi_{g,1/2}}$ band. Since the majority of the lines identified in that wavelength range are attributed to $\mathrm{CO_2^+}$, we measure the band intensity by fitting a continuum on both sides of the band, subtracting it, and then summing the flux over the entire width of the band. As mentioned before, all spectra have been subtracted of the dust continuum. The extra continuum subtraction done here is to remove any residual continuum component that might be left locally. For $\mathrm{N_{2}^+}$ and $\mathrm{CO^+}$, we use the (0,0) and (2,0) bands respectively. Since a non-negligible number of lines not attributed to those species are identified in the wavelength range of the aforementioned bands, we decided to use a different method to measure the intensity of the bands. We first fit and subtract a continuum for each band. In a second time, for each identified line of those two bands, we fit a gaussian and sum all lines of each band. When using the (390+580) setting, all three bands of interest are detected in a single set-up, allowing us to make the measurements from simultaneous observations.

The $\mathrm{N_{2}^+/CO^+}$ ratio is computed as:
\[ \mathrm{\frac{N_2^+}{CO^+} = \frac{g_{CO^+}}{g_{N_2^+}}}\frac{I_{N_2^+}}{I_{CO^+}}\]
where $\mathrm{I_{N_2^+}}$ and $\mathrm{I_{CO^+}}$ are the respective intensities of the $\mathrm{N_{2}^+}$ and $\mathrm{CO^+}$ bands, $\mathrm{g_{N_2^+}}$ is the excitation factor of $\mathrm{N_{2}^+}$ from \cite{Lutz1993} ($7\times 10^{-2}$ photons s$^{-1}$ mol$^{-1}$), and $\mathrm{g_{CO^+}}$ is the excitation factor of $\mathrm{CO^+}$ from \cite{Magnani1986} ($3.55\times 10^{-3}$ photons s$^{-1}$ mol$^{-1}$).

The $\mathrm{CO_{2}^+/CO^+}$ ratio is computed as:
\[ \mathrm{\frac{CO_2^+}{CO^+} = \frac{g_{CO^+}}{g_{CO_2^+}}}\frac{I_{CO_2^+}}{I_{CO^+}}\]
$\mathrm{I_{CO_2^+}}$ and $\mathrm{g_{CO_2^+}}$ are respectively the $\mathrm{CO_2^+}$ measured band intensity and excitation factor from \cite{Kim1999} ($4.958\times 10^{-4}$ photons s$^{-1}$ mol$^{-1}$). We measure a $\mathrm{N_{2}^+/CO^+}$ ratio of $0.06\pm0.01$, in very good agreement with \cite{Cochran2018}. Only very low upper limits of $1\times 10^{-5}$ to $1\times 10^{-4}$ for 122P/de Vico, C/1995 O1 (Hale-Bopp), and 153P/Ikeya-Zhang were derived by \cite{Cochran2000,Cochran2002}. Our ratio is higher than the ratio of 0.013 measured for 29P/Schwassmann-Wachmann \citep{Ivanova2016} but consistent with the ratio of 0.06 measured for C/2002 VQ94 (LINEAR) \citep{Korsun2014}.

For the $\mathrm{CO_{2}^+/CO^+}$ ratio we obtain a value of $1.1\pm0.3$. Simultaneous measurements of $\mathrm{CO^+}$ and $\mathrm{CO_2^+}$ ions in the coma of comets are scarce. It is then difficult to compare R2 to other comets. \cite{Feldman1997} used IUE and HST observations to constrain the $\mathrm{CO_2/CO}$ ratio in several comets. For comets for which both species could be detected, they found ratios ranging between 0.67 and 1.69. For two other comets, they could only derive lower limits of 4.6 and 1.8. However, we cannot infer directly the $\mathrm{CO_2/CO}$ ratio in the coma of R2 from the $\mathrm{CO_{2}^+/CO^+}$ ratio. Indeed, $\mathrm{CO^+}$ can be produced both from the photoionization of CO and photodissociative ionization of $\mathrm{CO_2}$. Few studies comparing the efficiencies of both processes have been performed so far, but \cite{Huebner1980} suggested that the photodissociative ionization of $\mathrm{CO_2}$ could significantly contribute to the production of $\mathrm{CO^+}$ in the coma of comets, especially in the inner coma. The relative contribution of both mechanisms in the coma of R2 is difficult to assess and would require dedicated modeling. Consequently, we must be cautious while interpreting the $\mathrm{CO_{2}^+/CO^+}$ and $\mathrm{N_{2}^+/CO^+}$ abundance ratios measured in the coma of this comet. 

In their work, as $\mathrm{CO_2^+}$ was not detected \cite{Cochran2018} consider that the photoionization of CO is the main source of $\mathrm{CO^+}$, in order to infer a $\mathrm{N_{2}/CO}$ ratio of 0.06 from their measurement of the $\mathrm{N_{2}^+/CO^+}$ ratio. If we take into account that part of the $\mathrm{CO^+}$ detected in the coma of R2 can be produced by the photodissociative ionization of $\mathrm{CO_2}$ in addition to the photoionization of CO, the $\mathrm{N_{2}/CO}$ of 0.06 reported by \cite{Cochran2018} would only be a lower limit. 

Since we do not detect $\mathrm{H_2O^+}$, we attempt to compute an upper limit of the $\mathrm{H_2O^+}/\mathrm{CO^+}$ ratio. We consider the $\mathrm{H_2O^+}$ $\mathrm{\tilde{A}^2A_1-\tilde{X}^2B_1}$ (8-0) band, together with the efficiency factor from \cite{Lutz1993} ($4.2\times 10^{-3}$ photons s$^{-1}$ mol$^{-1}$). Since that band is spread over a large wavelength range, over which numerous other emission lines are detected, integrating the flux over the whole wavelength range of the band would not provide us with a correct upper limit for the $\mathrm{H_2O^+}$ flux. Instead, we define zones spanning 0.015 nm on each side of the center of each line of the band in the cometary atlas of \cite{Cochran2002} and integrate the flux over those areas to have a more realistic estimate of the maximum total flux of the  $\mathrm{H_2O^+}$ band. Using this technique, we measure an upper limit of the $\mathrm{H_2O^+}/ \mathrm{CO^+}$ ratio of 0.4. \cite{Lutz1993} reported $\mathrm{H_2O^+}/ \mathrm{CO^+}$ ratios varying between 0.6 and 5 for a set of five comets. This is higher than our upper limit for R2, supporting the apparent low water abundance of this comet. However, it is important to keep in mind that the \cite{Lutz1993} measurements have been obtained for comets observed closer to the Sun, where water sublimation is more efficient. \cite{Biver2018} derived an upper limit for the water production rate of R2, leading to an upper limit of the $\mathrm{H_2O}/ \mathrm{CO}$ ratio of about 0.11. This is consistent with our low upper limit for the $\mathrm{H_2O^+}/ \mathrm{CO^+}$ ratio.

As mentioned earlier, we do not detect $\mathrm{NH_2}$ emission lines in the coma of R2. Nevertheless, deriving a lower limit of the $\mathrm{N_2^+}$/$\mathrm{NH_2}$ ratio would provide an interesting constrain for the solar nebula nitrogen chemistry. We used the $\mathrm{NH_2}$ A(0,8,0) - X(0,0,0) band. In the case of R2, the A(0,10,0) - X(0,0,0) band region shown in Fig. \ref{PlotNH2} contains a large number of $\mathrm{CO^+}$ emissions, making the computation of an upper limit of the $\mathrm{NH_2}$ flux difficult. Similarly to what is done for $\mathrm{H_2O^+}$, we define zones spanning 0.015 nm on each side of the center of each of the brightest lines of the band between 626 and 642 nm and integrated the flux over those areas. We use the fluorescence efficiency from \cite{Kawakita2002} ($1.07\times 10^{-3}$ photons s$^{-1}$ mol$^{-1}$). For all the ratios computed before, we assume an identical dependence of the fluorescence efficiencies with the heliocentric distance as $r_h^{-2}$. However, in the case of $\mathrm{NH_2}$, \cite{Kawakita2002} determined that the fluorescence efficiency actually varies as $r_h^{-1.6}$ and this is what we used. We then estimate the lower limit of the $\mathrm{N_2^+/NH_2}$ ratio to be of about 0.4.

\subsection{The nitrogen isotopic ratio} 

Nitrogen is one of the main species in comets, because of numerous molecules containing this atom and currently detected in their coma (NH$_3$, N$_2$, HCN, NH$_2$CHO, HNCO, HNC, CH$_3$CN, HC$_3$N, C$_2$H$_5$NO$_2$, CH$_5$N, C$_2$H$_7$N). So far it has been possible to measure the $^{14}$N/$^{15}$N ratio from CN and NH$_2$ optical spectra in about 20 comets of various origins. The isotopic ratios measured for these two molecules are in agreement within the errorbars: the averaged $^{14}$N/$^{15}$N ratio from NH$_2$ for 18 comets is 135.7 $\pm$ 5.9 \citep{Shinnaka2016} and agrees with that in CN for 22 comets: 145.2 $\pm$ 5.6 \citep{Manfroid2009}. It was also measured in HCN (the most obvious parent of CN) for comets 17P/Holmes and C/1995 O1 Hale-Bopp, with values of 139 and 205, respectively \citep{Bockelee2008}. The $^{14}$N/$^{15}$N ratio does not seem to change from one comet to another or from the reservoir of origin. Moreover, long and short period comets present the same ratio \citep{Jehin2009}. It is different from the ratio measured in the Earth's atmosphere (272 for the N$_2$ molecules according to \cite{nier1950}) or in the presolar nebula (estimated to be 441 from the solar wind, \cite{marty2011}) or 450 from Jupiter's atmosphere \citep{fouchet2004}). The reader is referred to \cite{bockeleemorvan2015} for a detailed discussion about isotopic ratios. The discovery of R2 presenting numerous N$_2^+$ bright emission lines offered a unique opportunity to detect $^{14}$N$^{15}$N$^+$ emission lines and, consequently, to measure the $^{14}$N/$^{15}$N isotopic ratio directly in N$_2$, the main reservoir of nitrogen in the solar nebula.

The first step consists in modeling as accurately as possible the
N$_2^+$ spectrum. The brightest emission lines belong to the (0,0) band of the first negative group, i.e. the $B^2\Sigma_u^+\rightarrow X^2\Sigma_u^+$ transition with a bandhead at 391.4~nm. A complete modeling of this spectrum would imply a fluorescence model taking into account all the transitions in this system, with the other bands and, possibly, other electronic transitions, both in absorption (of the solar flux) and in emission. We have not (yet) developed such a model but it is possible to do an acceptable modeling by using Boltzmann distribution(s), with parameters fitted to the observed spectrum.

\begin{figure*}
\centering
\includegraphics[width=20cm]{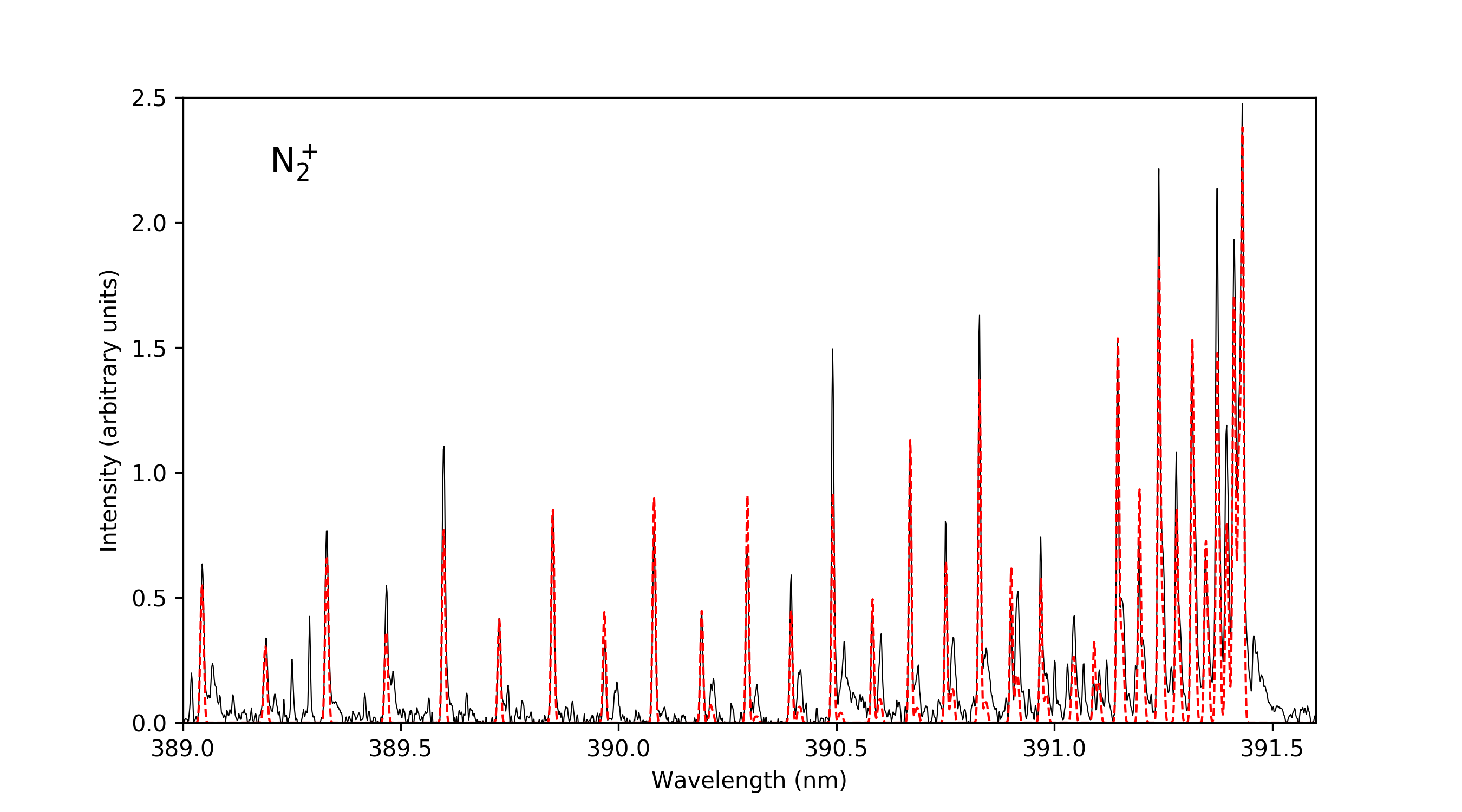}
\caption{Observed (black) and synthetic (red dashed) spectra of the N$_2^+$ (0,0) band of the first negative group in C/2016 R2.}
\label{PlotN2+_1}
\end{figure*}

To model the N$_2^+$ spectrum we first compute the energy levels from the constants published by \cite{wood1938} and \cite{lofthus1977} (their Table 72) and the line energies published by \cite{dick1978}. The Einstein coefficients for the spontaneous emission, A$_{v'v''}$ come from \cite{jain1967} and the H\"onl-London factors from \cite{herzberg1950} (only P and R lines exist for such a transition). From these data and the observed intensities of the cometary emission lines we compute a Boltzmann diagram to see if this spectrum follows a Boltzmann distribution. We find that the relative populations could be satisfactorily fitted by a double Boltzmann distribution based on two different rotational excitation temperatures, T$_{rot}\sim$45 K and 860 K (such a two-temperatures distribution is not unusual and has been observed for $\mathrm{C_2}$ for example, see \cite{Rousselot2012,Nelson2018} and references therein). From these temperatures we manage to fit the observed spectrum in a satisfying manner by adjusting the relative fraction of both populations. Of course such an approximation does not take into account solar absorption lines but, because of the numerous N$_2^+$ emission lines, the final fit is satisfactory. Fig. \ref{PlotN2+_1} presents the result in the range 389-391.6 nm.

The last step consists in computing a similar spectrum for the $^{14}$N$^{15}$N$^+$ isotopologue emission lines. It is done with the help of the line energies published by \cite{wood1938}. Only the emission lines belonging to the R branch are computed, because this paper does not contain the P branch data (or only very few), these emission lines being blended with the bright N$_2^+$ emission lines.

\begin{figure*}
\centering
\includegraphics[width=20cm]{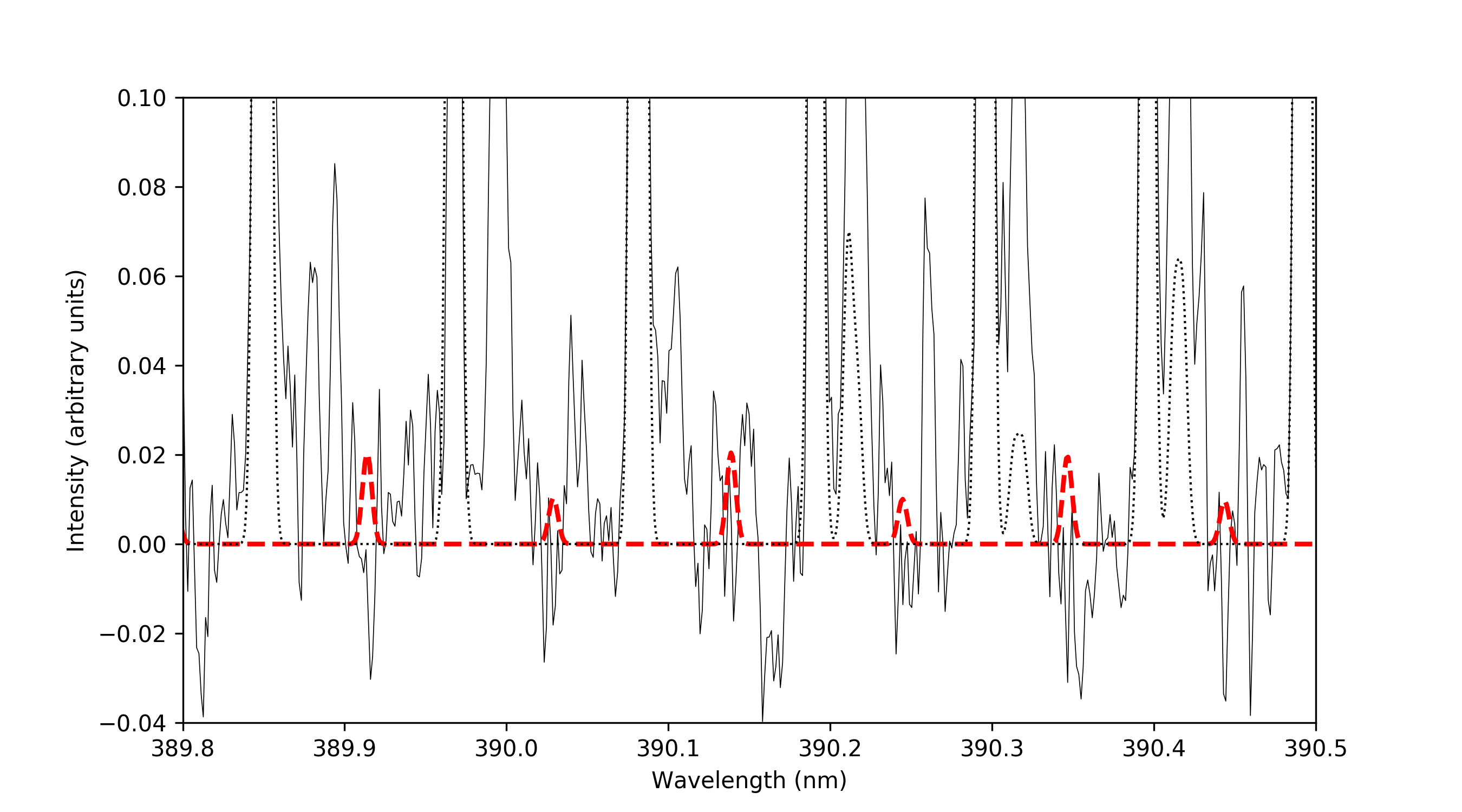}
\caption{Zoom of the observed spectrum (black) of C/2016 R2 and the
$^{14}$N$^{15}$N$^+$ synthetic spectrum computed for $^{14}$N/$^{15}$N=100 (dashed red curve). The dotted line represents the N$_2^+$ synthetic spectrum. In this part of the spectrum only the R lines (the brightest ones) of $^{14}$N$^{15}$N$^+$ are represented in the synthetic spectrum.}
\label{PlotN2+_2}
\end{figure*}

To compute the $^{14}$N$^{15}$N$^+$ emission spectrum we use similar rotational temperatures and ratios between them as fitted on N$_2^+$. Our search for the $^{14}$N$^{15}$N$^+$ is based on this spectrum. Fig. \ref{PlotN2+_2} presents the spectrum computed for $^{14}$N/$^{15}$N=100 (i.e. an intensity ratio of 50 with the N$_2^+$ emission lines). As it can be seen no $^{14}$N$^{15}$N$^+$ emission line is clearly identified. An estimate of the noise in the observational spectrum leads to about 0.02 (arbitrary) intensity units, to be compared to the maximum intensity of the $^{14}$N$^{15}$N$^+$ synthetic spectrum.

Our conclusion is that only a lower limit for the $^{14}$N/$^{15}$N ratio can be computed. This lower limit is roughly equal to 100.

\subsection{The dynamical history of R2} 

\begin{figure*}
   \centering
   \includegraphics[draft=false, width=0.8\textwidth]{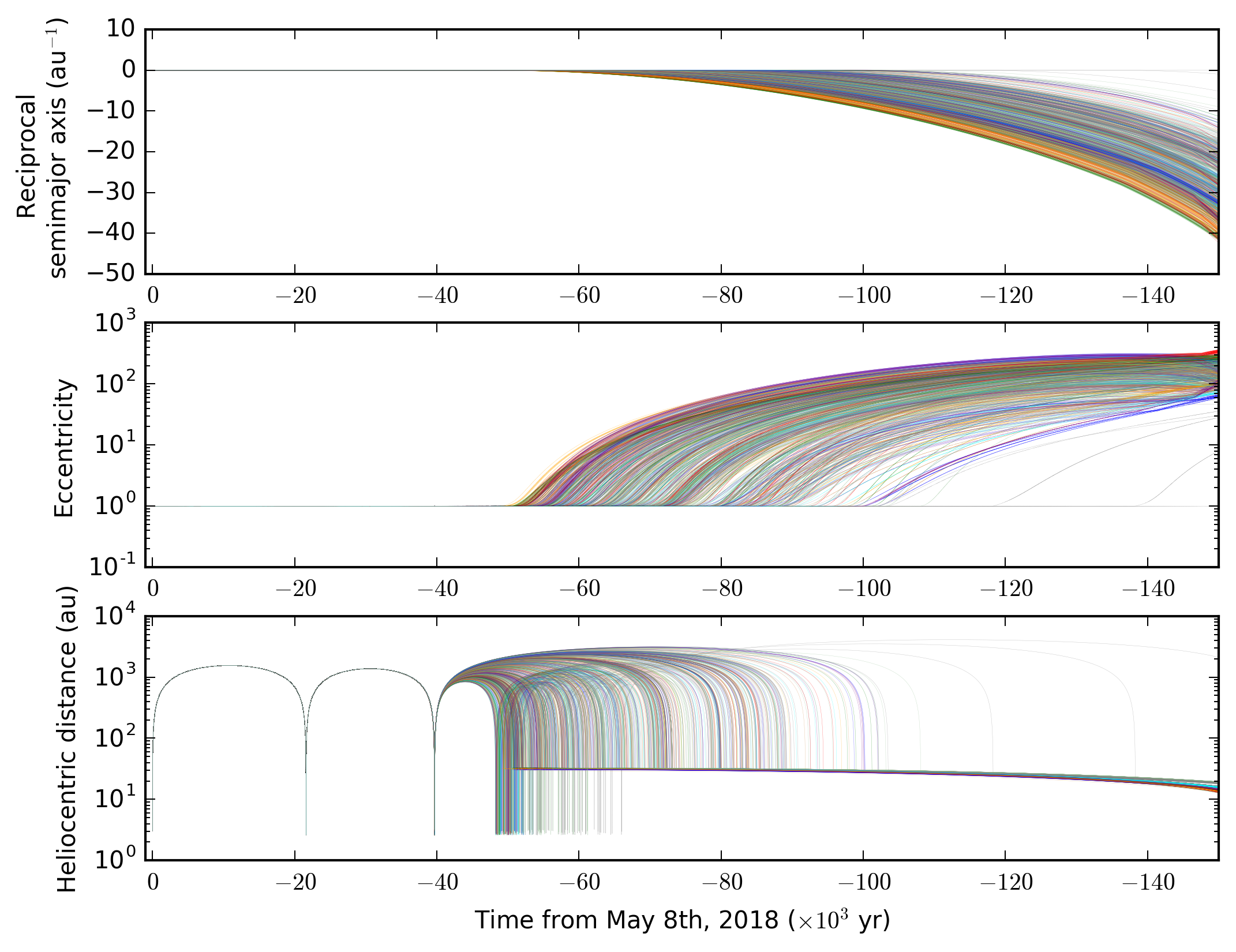}
       \caption{Orbital evolution of C/2016 R2 and 1000 clones over 1.5$\times10^{5}$ yr backward in time from last perihelion passage on May 8th, 2018. Different colors represent different 
       sets of 100 clones (see text for details). From the top to the bottom: reciprocal semimajor axis, eccentricity, and heliocentric distance. }
              \label{1000clones}%
    \end{figure*}

We outlined several times that R2 has a very unusual composition, showing strong emission lines of $\mathrm{N_2^+}$, $\mathrm{CO^+}$, and $\mathrm{CO_2^+}$, but no detected water or any of its dissociation product. In order to try to reach a better understanding of the origin of this comet, we decided to have a closer look at its dynamical history.

As we mentioned before, R2 is a nearly-isotropic comet originating from the Oort cloud, belonging to the subclass of returning objects, i.e., it is not the first time that this object has visited the inner region of the Solar System \citep{levison1996,luke2015}.
According to \cite{fernandez2005}, the computation of the original semi-major axis, $a_{org}$, and the orbital energy, $\chi_{org}$, is of great interest for long period comets since these parameters are vital to assess the place an object comes from and to determine its dynamical age, i.e., the average number of revolutions it has performed in the inner planetary region. This is particularly interesting in the case of R2, given its peculiar chemical composition.

By definition, the original orbit of a comet is the orbit a comet had before entering the planetary region, with respect to the barycenter of the Solar System (we refer the reader to \cite{fernandez2005} for further details about the computation of original orbits). R2 has an original reciprocal semi-major axis of $1/a_{org}=0.00129$ au$^{-1}$ (cf. the IAU/MPC Data Base\footnote{\url{https://www.minorplanetcenter.net/db_search}}). From observations and theoretical studies, \cite{fernandez2005} showed that a comet with such a reciprocal semi-major axis (or an equivalent orbital energy of $\chi_{org}=-0.00129$ au$^{-1}$) will perform about 10 revolutions before suffering an hyperbolic ejection ($\chi_{org}>0$). In order to assess the number of revolutions R2 has done, we decided to study its recent dynamical history through numerical simulations, as it has been done before by other authors for the study of long period comets (see, for example, \cite{hui2018u7c2}, \cite{hui2018k2}, for Jupiter family comets; \cite{pozuelos2018}, \cite{fernandez2015}, and for activated asteroids; \cite{novakovic2014}, and \cite{moreno2017j1}). 

In order to do that, we use numerical integrations in the heliocentric frame, the current time being set as the last perihelion passage, i.e., May 9th, 2018. We extend our integrations 3$\times10^{5}$ yr backward. We use the numerical package REBOUND \citep{rein2012}, with the integrator algorithm MERCURIUS, which is an hybrid integrator that combines a fast and unbiased symplectic Wisdom-Holman integrator, WHFAST \citep{rein2015whfast}, and a high accuracy non-symplectic integrator with adaptive time-stepping algorithm that handles close encounters, IAS15 \citep{rein2015ias15}. 
Because of the accumulation of errors during the numerical integrations, the dynamical evolution of a long period comet through successive passages by the planetary region is a stochastic process in which in every perihelion passage the comet meets a planetary configuration completely different from the previous one. To solve this, we perform a statistical study in which, in addition to the nominal orbit of the comet, the orbital evolution of a set of clones on orbits corresponding to the uncertainty is also followed. For this purpose, we generate 1000 clones based on the covariance matrix of the orbital elements \citep{cherni1998}.

In order to make the problem tractable with ordinary computational facilities, we split our 1000 clones in 10 sets of 100 clones. The set of orbital elements and the covariance matrix of the orbit of R2 are published together in the NASA/JPL Small-Body browser (JPL 20). The initial time-step is set to 5 days and the computed orbital evolution is stored every year for each clone. The Sun, the eight planets and Pluto are included in the simulation. Both galactic tide and gravitational effects of stars passing close to the Sun (which are the two main effects that contribute to eject comets from the Oort cloud to the inner solar system) are neglected. This assumption is valid because both effects are almost null for objects having semi-major axis well below $a\sim20000$ au \citep{souchay2010,luke2015}.
The action of non-gravitational forces might have a significant effect for long period comets approaching the Sun to less than a few tenths of au. However, the evaluation of these forces for long period comets is difficult since they have not been observed in a second apparition during which variations of the time of the perihelion passage could be detected. Thus, for simplicity, we do not include galactic tide, gravitational effects of stars passing close to the Sun, and non-gravitational forces in our simulations.

The orbital evolution of R2 and its 1000 clones in terms of reciprocal semi-major axis, $1/a$, eccentricity, $e$, and heliocentric distance in the past 1.5$\times10^{5}$ yr is shown in Fig. \ref{1000clones}. Our study shows that in a period between -(50-100) kyr 95$\%$ of the clones experience a dramatic change in their orbits, suffering hyperbolic ejections, i.e., $e>0$ and $1/a<0$. Therefore, we consider the upper limit of the time during which R2 has been orbiting in its current orbit to be of the order of 150 kyr. Moreover, our simulations show that 72.9$\%$ of the clones pass through perihelion only 3 times, 23.9$\%$ pass 4 times, 2.8$\%$ pass 5 times and 0.4$\%$ pass 6 times. In all cases, the perihelion distances 
are in the range of 2.6-3.2 au. These results show that R2 is in a dynamical state of evolution within the limits of "young" and  "middle-age" as defined in \cite{fernandez2005}.

This study shows that R2 had a relatively quiet recent dynamical history. As a consequence, it does not seem that its peculiar composition is related to its recent orbital evolution. The comet is not at its first passage through the planetary region, so it is not releasing unusual amounts of highly volatile species because it is approaching the Sun for the first time, but it has not been heavily processed due to repeated passages close to the Sun either. This indicates that, while the orbit of R2 is similar to many other Oort Cloud comets, its composition is intrinsically different.


\section{Discussion and Conclusions}

In this work, we present high resolution optical observations of comet C/2016 R2 (PanSTARRS) performed with the UVES instrument at the VLT, complemented by narrow-band imaging obtained with the TRAPPIST telescopes. R2 has a very peculiar chemical composition, with strong $\mathrm{N_2^+}$, $\mathrm{CO^+}$, and $\mathrm{CO_2^+}$ emission lines in the optical, faint CN, $\mathrm{C_3}$, and $\mathrm{C_2}$, and no detected OH, $\mathrm{OH^+}$ or $\mathrm{H_2O^+}$, suggesting a comet rich in $\mathrm{N_2}$ and CO but relatively poor in water. The comet also has clearly detected CH emissions. Table \ref{abundances} summarizes the ratios measured in the coma of R2. The non-detection of $\mathrm{NH_2}$ seems to suggest that most of the nitrogen content of R2 is concentrated in $\mathrm{N_2}$ with a lower limit of the $\mathrm{N_2^+}/\mathrm{NH_2}$ ratio of 0.4. Such a high $\mathrm{N_2^+}/\mathrm{NH_2}$ ratio might put strong constraints on the origin of R2 through adequate modeling \citep{Womack1992}. In some ways, R2 is similar to comets 29P/Schwassmann-Wachmann and C/2002 VQ94 (LINEAR), which are both distant active comets displaying clear $\mathrm{N_2^+}$ and $\mathrm{CO^+}$ emission lines.

\begin{table}
\caption{Ratios measured in the coma of R2}
\label{abundances}
\centering
\begin{tabular}{ll}
\hline\hline
Species &  Ratio\\
\hline
$\mathrm{N_2^+/CO^+}$ & $0.06\pm0.01$\\
$\mathrm{CO_2^+/CO^+}$ & $1.1\pm0.3$\\
$\mathrm{H_2O^+/CO^+}$ & $<0.4$\\
$\mathrm{N_2^+/NH_2}$ & $>0.4$\\
$log[Af\rho/Q(\mathrm{CN})]$ & $-21.78\pm0.14$ \\
\hline
\end{tabular}
\end{table}

Because we detect very strong $\mathrm{N_2^+}$ bands, we attempted to measure the $^{14}$N/$^{15}$N isotopic ratio from $\mathrm{N_2^+}$ directly for the first time. We could derive a lower limit of the $^{14}$N/$^{15}$N isotopic ratio of about 100, consistent with the values measured for comets from CN and $\mathrm{NH_2}$ \citep{bockeleemorvan2015}.

We determine an upper limit of about 0.4 for the $\mathrm{H_{2}O^+/CO^+}$ ratio, indicating a low water abundance in the coma of the comet. This is in agreement with our measurement of the ratio between the green oxygen forbidden line at 557.73 nm and the red doublet at 630.03 and 636.37 nm. This quantity can be linked to the mixing ratios of $\mathrm{CO/H_2O}$ and $\mathrm{CO_2/H_2O}$. We measure a G/R ratio of 0.23, consistent with what has been measured by \cite{Decock2013} for comets observed at large distances from the Sun ($>$3.0 au), when water sublimation is less efficient and $\mathrm{CO}$ and $\mathrm{CO_2}$ are significantly contributing to the cometary activity.

For the first time to our knowledge, we clearly detected the [NI] forbidden lines at 519.79 and 520.02 nm in the coma of a comet. These lines are the result of the decay of nitrogen in a metastable state most probably produced in this case by the dissociative electron recombination of $\mathrm{N_2^+}$, electron impact dissociative ionization of $\mathrm{N_2}$, and/or the electron impact dissociation of $\mathrm{N_2}$. We conducted a search in our database of high resolution spectra of a dozen comets observed with UVES over the past 15 years. We do not detect the [NI] lines in any of those comets, among which some have been observed under circumstances similar to those of R2. The unusually high abundance of molecular nitrogen in the coma of R2 might then explain why we detect the [NI] lines in this particular comet and not in others.

The presence of highly volatile species such as $\mathrm{N_2}$ or $\mathrm{CO}$ in unusually high amounts while other species like water, CN, or $\mathrm{NH_2}$ are depleted is difficult to explain. One could wonder whether this comet has undergone only very minor processing since its formation, for example. However, our analysis of its dynamical history shows that R2 is a dynamically middle-aged comet coming from the Oort cloud, having crossed the planetary region at least 3 times in the past, with perihelion distances of around 2.5 to 3.0 au. Therefore the peculiar composition of its coma cannot be explained by the fact that it is releasing highly volatile gases far away from the Sun on its first perihelion passage, even though it probably did not undergo strong thermal processing due to a close encounter with the Sun.

Another hypothesis is that the origin of its peculiarity might then be intrinsic and linked to the conditions prevailing in the solar nebula at the time of its formation, to the distance from the Sun, or the time at which it formed. Since we detect significant $\mathrm{CO_2^+}$ emission lines, we cannot derive the $\mathrm{N_{2}/CO}$ ratio from our measurement of the $\mathrm{N_{2}^+/CO^+}$. However, we can set a lower limit of 0.06. This value is about 10 times higher than the one measured in comet 67P, which was interpreted as an indication of a lower limit of 24 K for the temperature experienced by the grains when they agglomerated to form 67P \citep{Rubin2015}. This interpretation is based on the results of laboratory experiments published by \cite{BarNun2007} and their estimation of the $\mathrm{N_{2}/CO}$ ratio in the solar nebula of 0.22. This estimation was multiplied by their average depletion factor for $\mathrm{N_{2}/CO}$ measured at three different temperatures (24, 27 and 30 K), which lead to a present $\mathrm{N_{2}/CO}$ in comets of $6.6\times10^{-3}$, very close to the value of $(5.7\pm0.66)\times10^{-3}$ measured in 67P. For R2 we estimate this ratio one order of magnitude above, with a large amount of $\mathrm{N_{2}}$ and CO, indicating that, based on the measured efficiency of gas trapping in water ice, the temperature experienced by the grains when they agglomerated was lower than 24 K. It should, nevertheless, be pointed out that different attempts to measure $\mathrm{N_{2}/CO}$ from $\mathrm{N_{2}^+/CO^+}$ with ground-based facilities \citep{Cochran2000,Cochran2002} in different comets (153P/Ikeya-Zhang, C/1995 O1 Hale-Bopp and 122P/1995 S1 de Vico) provided upper limits for the $\mathrm{N_{2}/CO}$ ratio to be one order of magnitude lower than the ratio measured in the coma of 67P. Because these attempts of measuring this ratio are based on a very different observational technique compared to the in-situ ROSINA measurements an observational bias cannot be excluded. These results, nevertheless, seem to indicate that both 67P and R2 are enriched in $\mathrm{N_{2}}$ compared to most comets, the last showing the strongest enrichement in $\mathrm{N_{2}}$. Because of the very different dynamical properties of these two comets it is probable that the lower $\mathrm{N_{2}/CO}$ ratio observed in comet 67P is also due to the effect of repeated solar heating on its nucleus, that led to a loss of the original $\mathrm{N_{2}}$. At the end we can only speculate that the temperature of the formation region of R2 was lower than the one of 67P but it is hard to be more accurate on this point. 

Alternatively, \cite{Biver2018} suggest that R2 could be a collisional fragment of a Kuiper Belt object (KBO). Indeed, some Kuiper Belt objects such as Pluto or Eris have $\mathrm{N_2}$-rich surfaces. Large KBOs are differentiated and the radiogenic heating they undergo would release highly volatile gases, which will then re-condense in the outer layers, enriching those layers in volatiles such as $\mathrm{N_2}$ or CO. Another clue that could point toward R2 being a KBO fragment is the strong CH emission we detected, which could point to a relatively high methane content. KBOs like Pluto and Eris have been shown to have a surface rich in methane. However, even though it is dynamically possible, the likelihood of a KBO fragment ending-up on an Oort Cloud orbit should be investigated. As a conclusion, more modelling is needed to understand the origin of the peculiar chemical composition of R2, both in terms of its place and conditions of formation and in terms of its evolutionary path.

\begin{acknowledgements}
Based on observations made with ESO Telescopes at the La Silla Paranal Observatory under program 2100.C-5035(A). TRAPPIST-South is a project funded by the Belgian Fonds (National) de la Recherche Scientifique (F.R.S.-FNRS) under grant FRFC 2.5.594.09.F. TRAPPIST-North is a project funded by the University of Liège, and performed in collaboration with Cadi Ayyad University of Marrakesh. CO is an ESO fellow. DH, EJ and MG are FNRS Senior Research Associates. YM acknowledges the support of Erasmus+ International Credit Mobility. FJP is supported by a Marie Curie CO-FUND fellowship, co-founded by the University of Liège and the European Union. We thank Julio Fernandez for his valuable discussion. Simulations in this paper made use of the REBOUND code which can be downloaded freely at http://github.com/hannorein/rebound. We thank Hanno Rein and Daniel Tamayo for their help using REBOUND. We thank R. Thomas and B. Häußler for performing the observations.    
\end{acknowledgements}


\bibliographystyle{aa}

\begin{thebibliography}{81}
\expandafter\ifx\csname natexlab\endcsname\relax\def\natexlab#1{#1}\fi

\bibitem[{{A'Hearn} {et~al.}(1995){A'Hearn}, {Millis}, {Schleicher}, {Osip}, \&
  {Birch}}]{AHearn1995}
{A'Hearn}, M.~F., {Millis}, R.~C., {Schleicher}, D.~O., {Osip}, D.~J., \&
  {Birch}, P.~V. 1995, \icarus, 118, 223

\bibitem[{{Aller} \& {Walker}(1970)}]{Aller1970}
{Aller}, L.~H. \& {Walker}, M.~F. 1970, \apj, 161, 917

\bibitem[{{Bar-Nun} {et~al.}(2007){Bar-Nun}, {Notesco}, \& {Owen}}]{BarNun2007}
{Bar-Nun}, A., {Notesco}, G., \& {Owen}, T. 2007, \icarus, 190, 655

\bibitem[{{Bieler} {et~al.}(2015){Bieler}, {Altwegg}, {Balsiger}, {Bar-Nun},
  {Berthelier}, {Bochsler}, {Briois}, {Calmonte}, {Combi}, {de Keyser}, {van
  Dishoeck}, {Fiethe}, {Fuselier}, {Gasc}, {Gombosi}, {Hansen}, {H{\"a}ssig},
  {J{\"a}ckel}, {Kopp}, {Korth}, {Le Roy}, {Mall}, {Maggiolo}, {Marty},
  {Mousis}, {Owen}, {R{\`e}me}, {Rubin}, {S{\'e}mon}, {Tzou}, {Waite}, {Walsh},
  \& {Wurz}}]{Bieler2015}
{Bieler}, A., {Altwegg}, K., {Balsiger}, H., {et~al.} 2015, \nat, 526, 678

\bibitem[{{Biver} {et~al.}(2018){Biver}, {Bockel{\'e}e-Morvan}, {Paubert},
  {Moreno}, {Crovisier}, {Boissier}, {Bertrand}, {Boussier}, {Kugel}, {McKay},
  {Dello Russo}, \& {DiSanti}}]{Biver2018}
{Biver}, N., {Bockel{\'e}e-Morvan}, D., {Paubert}, G., {et~al.} 2018, ArXiv
  e-prints [\eprint[arXiv]{1809.08086}]

\bibitem[{{Bockel{\'e}e-Morvan} {et~al.}(2008){Bockel{\'e}e-Morvan}, {Biver},
  {Jehin}, {Cochran}, {Wiesemeyer}, {Manfroid}, {Hutsem{\'e}kers}, {Arpigny},
  {Boissier}, {Cochran}, {Colom}, {Crovisier}, {Milutinovic}, {Moreno},
  {Prochaska}, {Ramirez}, {Schulz}, \& {Zucconi}}]{Bockelee2008}
{Bockel{\'e}e-Morvan}, D., {Biver}, N., {Jehin}, E., {et~al.} 2008, \apjl, 679,
  L49

\bibitem[{{Bockel{\'e}e-Morvan} {et~al.}(2015){Bockel{\'e}e-Morvan},
  {Calmonte}, {Charnley}, {Duprat}, {Engrand}, {Gicquel}, {H{\"a}ssig},
  {Jehin}, {Kawakita}, {Marty}, {Milam}, {Morse}, {Rousselot}, {Sheridan}, \&
  {Wirstr{\"o}m}}]{bockeleemorvan2015}
{Bockel{\'e}e-Morvan}, D., {Calmonte}, U., {Charnley}, S., {et~al.} 2015, SSR,
  197, 47

\bibitem[{{Chernitsov} {et~al.}(1998){Chernitsov}, {Baturin}, \&
  {Tamarov}}]{cherni1998}
{Chernitsov}, A.~M., {Baturin}, A.~P., \& {Tamarov}, V.~A. 1998, Solar System
  Research, 32, 405

\bibitem[{{Cochran}(2002)}]{Cochran2002}
{Cochran}, A.~L. 2002, \apjl, 576, L165

\bibitem[{{Cochran} {et~al.}(2000){Cochran}, {Cochran}, \&
  {Barker}}]{Cochran2000}
{Cochran}, A.~L., {Cochran}, W.~D., \& {Barker}, E.~S. 2000, \icarus, 146, 583

\bibitem[{{Cochran} \& {McKay}(2018)}]{Cochran2018}
{Cochran}, A.~L. \& {McKay}, A.~J. 2018, \apjl, 854, L10

\bibitem[{{Cravens} \& {Green}(1978)}]{Cravens1978}
{Cravens}, T.~E. \& {Green}, A.~E.~S. 1978, \icarus, 33, 612

\bibitem[{{de La Baume Pluvinel} \& {Baldet}(1911)}]{Pluvinel1911}
{de La Baume Pluvinel}, A. \& {Baldet}, F. 1911, \apj, 34, 89

\bibitem[{{de Val-Borro} {et~al.}(2018){de Val-Borro}, {Milam}, {Cordiner},
  {Charnley}, {Villanueva}, \& {Kuan}}]{deValBorro2018}
{de Val-Borro}, M., {Milam}, S.~N., {Cordiner}, M.~A., {et~al.} 2018, The
  Astronomer's Telegram, 11254

\bibitem[{{Decock} {et~al.}(2013){Decock}, {Jehin}, {Hutsem{\'e}kers}, \&
  {Manfroid}}]{Decock2013}
{Decock}, A., {Jehin}, E., {Hutsem{\'e}kers}, D., \& {Manfroid}, J. 2013, \aap,
  555, A34

\bibitem[{{Decock} {et~al.}(2015){Decock}, {Jehin}, {Rousselot},
  {Hutsem{\'e}kers}, {Manfroid}, {Raghuram}, {Bhardwaj}, \&
  {Hubert}}]{Decock2015}
{Decock}, A., {Jehin}, E., {Rousselot}, P., {et~al.} 2015, \aap, 573, A1

\bibitem[{{Dick} {et~al.}(1978){Dick}, {Benesch}, {Crosswhite}, {Tilford},
  {Gottscho}, \& {Field}}]{dick1978}
{Dick}, K.~A., {Benesch}, W., {Crosswhite}, H.~M., {et~al.} 1978, Journal of
  Molecular Spectroscopy, 69, 95

\bibitem[{{Dones} {et~al.}(2015){Dones}, {Brasser}, {Kaib}, \&
  {Rickman}}]{luke2015}
{Dones}, L., {Brasser}, R., {Kaib}, N., \& {Rickman}, H. 2015, \ssr, 197, 191

\bibitem[{{Dopita} {et~al.}(1976){Dopita}, {Mason}, \& {Robb}}]{Dopita1976}
{Dopita}, M.~A., {Mason}, D.~J., \& {Robb}, W.~D. 1976, \apj, 207, 102

\bibitem[{{Farnham} {et~al.}(2000){Farnham}, {Schleicher}, \&
  {A'Hearn}}]{Farnham2000}
{Farnham}, T.~L., {Schleicher}, D.~G., \& {A'Hearn}, M.~F. 2000, \icarus, 147,
  180

\bibitem[{{Feldman} {et~al.}(1997){Feldman}, {Festou}, {Tozzi}, \&
  {Weaver}}]{Feldman1997}
{Feldman}, P.~D., {Festou}, M.~C., {Tozzi}, P., \& {Weaver}, H.~A. 1997, \apj,
  475, 829

\bibitem[{Fern{\'a}ndez(2005)}]{fernandez2005}
Fern{\'a}ndez, J.~A. 2005, in {Astrophysics and Space Science Library}, Vol.
  328 (Springer, Dordrecht)

\bibitem[{{Fern{\'a}ndez} \& {Sosa}(2015)}]{fernandez2015}
{Fern{\'a}ndez}, J.~A. \& {Sosa}, A. 2015, \planss, 118, 14

\bibitem[{{Festou} \& {Feldman}(1981)}]{Festou1981}
{Festou}, M. \& {Feldman}, P.~D. 1981, \aap, 103, 154

\bibitem[{{Fouchet} {et~al.}(2004){Fouchet}, {Irwin}, {Parrish}, {Calcutt},
  {Taylor}, {Nixon}, \& {Owen}}]{fouchet2004}
{Fouchet}, T., {Irwin}, P.~G.~J., {Parrish}, P., {et~al.} 2004, Icarus, 172, 50

\bibitem[{{Greenstein}(1962)}]{Greenstein1962}
{Greenstein}, J.~L. 1962, \apj, 136, 688

\bibitem[{{Haser}(1957)}]{Haser1957}
{Haser}, L. 1957, Bulletin de la Societe Royale des Sciences de Liege, 43, 740

\bibitem[{{Herzberg}(1950)}]{herzberg1950}
{Herzberg}, G. 1950, {Molecular spectra and molecular structure. Vol.1: Spectra
  of diatomic molecules}

\bibitem[{{Hibbert} {et~al.}(1993){Hibbert}, {Biemont}, {Godefroid}, \&
  {Vaeck}}]{Hibbert1993}
{Hibbert}, A., {Biemont}, E., {Godefroid}, M., \& {Vaeck}, N. 1993, \aaps, 99,
  179

\bibitem[{{Huebner} \& {Giguere}(1980)}]{Huebner1980}
{Huebner}, W.~F. \& {Giguere}, P.~T. 1980, \apj, 238, 753

\bibitem[{{Hui}(2018)}]{hui2018u7c2}
{Hui}, M.-T. 2018, \aj, 156, 73

\bibitem[{{Hui} {et~al.}(2018){Hui}, {Jewitt}, \& {Clark}}]{hui2018k2}
{Hui}, M.-T., {Jewitt}, D., \& {Clark}, D. 2018, \aj, 155, 25

\bibitem[{{Iro} {et~al.}(2003){Iro}, {Gautier}, {Hersant},
  {Bockel{\'e}e-Morvan}, \& {Lunine}}]{Iro2003}
{Iro}, N., {Gautier}, D., {Hersant}, F., {Bockel{\'e}e-Morvan}, D., \&
  {Lunine}, J.~I. 2003, \icarus, 161, 511

\bibitem[{{Ivanova} {et~al.}(2016){Ivanova}, {Luk`yanyk}, {Kiselev},
  {Afanasiev}, {Picazzio}, {Cavichia}, {de Almeida}, \&
  {Andrievsky}}]{Ivanova2016}
{Ivanova}, O.~V., {Luk`yanyk}, I.~V., {Kiselev}, N.~N., {et~al.} 2016, \planss,
  121, 10

\bibitem[{{Ivanova} {et~al.}(2018){Ivanova}, {Picazzio}, {Luk'yanyk},
  {Cavichia}, \& {Andrievsky}}]{ivanova2018}
{Ivanova}, O.~V., {Picazzio}, E., {Luk'yanyk}, I.~V., {Cavichia}, O., \&
  {Andrievsky}, S.~M. 2018, PASS, 157, 34

\bibitem[{{Jain} \& {Sahni}(1967)}]{jain1967}
{Jain}, D.~C. \& {Sahni}, R.~C. 1967, Int. J. Quantum Chem., 1, 721

\bibitem[{{Jehin} {et~al.}(2011){Jehin}, {Gillon}, {Queloz}, {Magain},
  {Manfroid}, {Chantry}, {Lendl}, {Hutsem{\'e}kers}, \& {Udry}}]{Jehin2011}
{Jehin}, E., {Gillon}, M., {Queloz}, D., {et~al.} 2011, The Messenger, 145, 2

\bibitem[{{Jehin} {et~al.}(2009){Jehin}, {Manfroid}, {Hutsem{\'e}kers},
  {Arpigny}, \& {Zucconi}}]{Jehin2009}
{Jehin}, E., {Manfroid}, J., {Hutsem{\'e}kers}, D., {Arpigny}, C., \&
  {Zucconi}, J.-M. 2009, Earth Moon and Planets, 105, 167

\bibitem[{{Kawakita} \& {Watanabe}(2002)}]{Kawakita2002}
{Kawakita}, H. \& {Watanabe}, J.-i. 2002, \apjl, 572, L177

\bibitem[{{Kim}(1999)}]{Kim1999}
{Kim}, S.~J. 1999, Earth, Planets, and Space, 51, 139

\bibitem[{{Korsun} {et~al.}(2008){Korsun}, {Ivanova}, \&
  {Afanasiev}}]{Korsun2008}
{Korsun}, P.~P., {Ivanova}, O.~V., \& {Afanasiev}, V.~L. 2008, \icarus, 198,
  465

\bibitem[{{Korsun} {et~al.}(2014){Korsun}, {Rousselot}, {Kulyk}, {Afanasiev},
  \& {Ivanova}}]{Korsun2014}
{Korsun}, P.~P., {Rousselot}, P., {Kulyk}, I.~V., {Afanasiev}, V.~L., \&
  {Ivanova}, O.~V. 2014, \icarus, 232, 88

\bibitem[{{Kumar} \& {Shashikiran}(2018)}]{Kumar2018}
{Kumar}, V. \& {Shashikiran}, G. 2018, EPSC Abstracts, European Planetary
  Science Congress 2018, 12

\bibitem[{{Levison}(1996)}]{levison1996}
{Levison}, H.~F. 1996, in {Completing the Inventory of the Solar System}, ed.
  T.~{Rettig} \& J.~M. {Hahn}, Vol. 107, 173--191

\bibitem[{{Lofthus} \& {Krupenie}(1977)}]{lofthus1977}
{Lofthus}, A. \& {Krupenie}, P.~H. 1977, Journal of Physical and Chemical
  Reference Data, 6, 113

\bibitem[{{Lutz} {et~al.}(1993){Lutz}, {Womack}, \& {Wagner}}]{Lutz1993}
{Lutz}, B.~L., {Womack}, M., \& {Wagner}, R.~M. 1993, \apj, 407, 402

\bibitem[{{Magnani} \& {A'Hearn}(1986)}]{Magnani1986}
{Magnani}, L. \& {A'Hearn}, M.~F. 1986, \apj, 302, 477

\bibitem[{{Manfroid} {et~al.}(2005){Manfroid}, {Jehin}, {Hutsem{\'e}kers},
  {Cochran}, {Zucconi}, {Arpigny}, {Schulz}, \& {St{\"u}we}}]{Manfroid2005}
{Manfroid}, J., {Jehin}, E., {Hutsem{\'e}kers}, D., {et~al.} 2005, \aap, 432,
  L5

\bibitem[{{Manfroid} {et~al.}(2009){Manfroid}, {Jehin}, {Hutsem{\'e}kers},
  {Cochran}, {Zucconi}, {Arpigny}, {Schulz}, {St{\"u}we}, \&
  {Ilyin}}]{Manfroid2009}
{Manfroid}, J., {Jehin}, E., {Hutsem{\'e}kers}, D., {et~al.} 2009, \aap, 503,
  613

\bibitem[{{Marty} {et~al.}(2011){Marty}, {Chaussidon}, {Wiens}, {Jurewicz}, \&
  {Burnett}}]{marty2011}
{Marty}, B., {Chaussidon}, M., {Wiens}, R.~C., {Jurewicz}, A.~J.~G., \&
  {Burnett}, D.~S. 2011, Science, 332, 1533

\bibitem[{{McKay} {et~al.}(2012){McKay}, {Chanover}, {Morgenthaler}, {Cochran},
  {Harris}, \& {Russo}}]{McKay2012}
{McKay}, A.~J., {Chanover}, N.~J., {Morgenthaler}, J.~P., {et~al.} 2012,
  \icarus, 220, 277

\bibitem[{{McKay} {et~al.}(2013){McKay}, {Chanover}, {Morgenthaler}, {Cochran},
  {Harris}, \& {Russo}}]{McKay2013}
{McKay}, A.~J., {Chanover}, N.~J., {Morgenthaler}, J.~P., {et~al.} 2013,
  \icarus, 222, 684

\bibitem[{{McKay} {et~al.}(2015){McKay}, {Cochran}, {DiSanti}, {Villanueva},
  {Russo}, {Vervack}, {Morgenthaler}, {Harris}, \& {Chanover}}]{McKay2015}
{McKay}, A.~J., {Cochran}, A.~L., {DiSanti}, M.~A., {et~al.} 2015, \icarus,
  250, 504

\bibitem[{{Moreno} {et~al.}(2017){Moreno}, {Pozuelos}, {Novakovi{\'c}},
  {Licandro}, {Cabrera-Lavers}, {Bolin}, {Jedicke}, {Gladman}, {Bannister},
  {Gwyn}, {Vere\v{s}}, {Chambers}, {Chastel}, {Denneau}, {Flewelling}, {Huber},
  {Schunov{\'a}-Lilly}, {Magnier}, {Wainscoat}, {Waters}, {Weryk},
  {Farnocchia}, \& {Micheli}}]{moreno2017j1}
{Moreno}, F., {Pozuelos}, F.~J., {Novakovi{\'c}}, B., {et~al.} 2017, \apjl,
  837, L3

\bibitem[{{Mousis} {et~al.}(2012){Mousis}, {Guilbert-Lepoutre}, {Lunine},
  {Cochran}, {Waite}, {Petit}, \& {Rousselot}}]{Mousis2012}
{Mousis}, O., {Guilbert-Lepoutre}, A., {Lunine}, J.~I., {et~al.} 2012, \apj,
  757, 146

\bibitem[{{Nelson} {et~al.}(2018){Nelson}, {Cochran}, \&
  {Western}}]{Nelson2018}
{Nelson}, T., {Cochran}, A.~L., \& {Western}, C. 2018, ArXiv e-prints
  [\eprint[arXiv]{1809.07715}]

\bibitem[{{Nier}(1950)}]{nier1950}
{Nier}, A.~O. 1950, Physical Review, 77, 789

\bibitem[{{Novakovi{\'c}} {et~al.}(2014){Novakovi{\'c}}, {Hsieh}, {Cellino},
  {Micheli}, \& {Pedani}}]{novakovic2014}
{Novakovi{\'c}}, B., {Hsieh}, H.~H., {Cellino}, A., {Micheli}, M., \& {Pedani},
  M. 2014, \icarus, 231, 300

\bibitem[{{Oliversen} {et~al.}(2002){Oliversen}, {Doane}, {Scherb}, {Harris},
  \& {Morgenthaler}}]{Oliversen2002}
{Oliversen}, R.~J., {Doane}, N., {Scherb}, F., {Harris}, W.~M., \&
  {Morgenthaler}, J.~P. 2002, \apj, 581, 770

\bibitem[{{Ootsubo} {et~al.}(2012){Ootsubo}, {Kawakita}, {Hamada}, {Kobayashi},
  {Yamaguchi}, {Usui}, {Nakagawa}, {Ueno}, {Ishiguro}, {Sekiguchi}, {Watanabe},
  {Sakon}, {Shimonishi}, \& {Onaka}}]{Ootsubo2012}
{Ootsubo}, T., {Kawakita}, H., {Hamada}, S., {et~al.} 2012, \apj, 752, 15

\bibitem[{{Opitom} {et~al.}(2015){Opitom}, {Jehin}, {Manfroid},
  {Hutsem{\'e}kers}, {Gillon}, \& {Magain}}]{Opitom2015}
{Opitom}, C., {Jehin}, E., {Manfroid}, J., {et~al.} 2015, \aap, 574, A38

\bibitem[{{Owen} \& {Bar-Nun}(1995)}]{Owen1995}
{Owen}, T. \& {Bar-Nun}, A. 1995, \icarus, 116, 215

\bibitem[{{Pozuelos} {et~al.}(2018){Pozuelos}, {Jehin}, {Moulane}, {Opitom},
  {Manfroid}, {Benkhaldoun}, \& {Gillon}}]{pozuelos2018}
{Pozuelos}, F.~J., {Jehin}, E., {Moulane}, Y., {et~al.} 2018, \aap, 615, A154

\bibitem[{{Rees} \& {Romick}(1985)}]{Rees1985}
{Rees}, M.~H. \& {Romick}, G.~J. 1985, \jgr, 90, 9871

\bibitem[{{Rein} \& {Liu}(2012)}]{rein2012}
{Rein}, H. \& {Liu}, S.-F. 2012, \aap, 537, A128

\bibitem[{{Rein} \& {Spiegel}(2015)}]{rein2015ias15}
{Rein}, H. \& {Spiegel}, D.~S. 2015, \mnras, 446, 1424

\bibitem[{{Rein} \& {Tamayo}(2015)}]{rein2015whfast}
{Rein}, H. \& {Tamayo}, D. 2015, \mnras, 452, 376

\bibitem[{{Rousselot} {et~al.}(2012){Rousselot}, {Jehin}, {Manfroid}, \&
  {Hutsem{\'e}kers}}]{Rousselot2012}
{Rousselot}, P., {Jehin}, E., {Manfroid}, J., \& {Hutsem{\'e}kers}, D. 2012,
  \aap, 545, A24

\bibitem[{{Rubin} {et~al.}(2015){Rubin}, {Altwegg}, {Balsiger}, {Bar-Nun},
  {Berthelier}, {Bieler}, {Bochsler}, {Briois}, {Calmonte}, {Combi}, {De
  Keyser}, {Dhooghe}, {Eberhardt}, {Fiethe}, {Fuselier}, {Gasc}, {Gombosi},
  {Hansen}, {H{\"a}ssig}, {J{\"a}ckel}, {Kopp}, {Korth}, {Le Roy}, {Mall},
  {Marty}, {Mousis}, {Owen}, {R{\`e}me}, {S{\'e}mon}, {Tzou}, {Waite}, \&
  {Wurz}}]{Rubin2015}
{Rubin}, M., {Altwegg}, K., {Balsiger}, H., {et~al.} 2015, Science, 348, 232

\bibitem[{{Sharpee} {et~al.}(2005){Sharpee}, {Slanger}, {Cosby}, \&
  {Huestis}}]{Sharpee2005}
{Sharpee}, B.~D., {Slanger}, T.~G., {Cosby}, P.~C., \& {Huestis}, D.~L. 2005,
  \grl, 32, L12106

\bibitem[{{Shinnaka} {et~al.}(2016){Shinnaka}, {Kawakita}, {Jehin}, {Decock},
  {Hutsem{\'e}kers}, {Manfroid}, \& {Arai}}]{Shinnaka2016}
{Shinnaka}, Y., {Kawakita}, H., {Jehin}, E., {et~al.} 2016, \mnras, 462, S195

\bibitem[{{Singh} {et~al.}(1991){Singh}, {D'Ealmeida}, \&
  {Huebner}}]{Singh1991}
{Singh}, P.~D., {D'Ealmeida}, A.~A., \& {Huebner}, W.~F. 1991, \icarus, 90, 74

\bibitem[{{Smette} {et~al.}(2015){Smette}, {Sana}, {Noll}, {Horst}, {Kausch},
  {Kimeswenger}, {Barden}, {Szyszka}, {Jones}, {Gallenne}, {Vinther},
  {Ballester}, \& {Taylor}}]{Smette2015}
{Smette}, A., {Sana}, H., {Noll}, S., {et~al.} 2015, \aap, 576, A77

\bibitem[{{Souchay} \& {Dvorak}(2010)}]{souchay2010}
{Souchay}, J. \& {Dvorak}, R. 2010, {Dynamics of Small Solar System Bodies and
  Exoplanets}

\bibitem[{{Weryk} \& {Wainscoat}(2016)}]{Weryk2016}
{Weryk}, R. \& {Wainscoat}, R. 2016, Central Bureau Electronic Telegrams, 4318

\bibitem[{{Wierzchos} \& {Womack}(2017)}]{Wierzchos2017}
{Wierzchos}, K. \& {Womack}, M. 2017, Central Bureau Electronic Telegrams, 4464

\bibitem[{{Wierzchos} \& {Womack}(2018)}]{Wierzchos2018}
{Wierzchos}, K. \& {Womack}, M. 2018, ArXiv e-prints
  [\eprint[arXiv]{1805.06918}]

\bibitem[{{Wiese} {et~al.}(1966){Wiese}, {Smith}, \& {Glennon}}]{Wiese1966}
{Wiese}, W.~L., {Smith}, M.~W., \& {Glennon}, B.~M. 1966, {Atomic transition
  probabilities. Vol.: Hydrogen through Neon. A critical data compilation}

\bibitem[{{Womack} {et~al.}(1992){Womack}, {Wyckoff}, \& {Ziurys}}]{Womack1992}
{Womack}, M., {Wyckoff}, S., \& {Ziurys}, L.~M. 1992, \apj, 401, 728

\bibitem[{{Wood} \& {Dieke}(1938)}]{wood1938}
{Wood}, R.~H. \& {Dieke}, G. 1938, JCP, 6, 734

\bibitem[{{Wyckoff} \& {Theobald}(1989)}]{Wyckoff1989}
{Wyckoff}, S. \& {Theobald}, J. 1989, Advances in Space Research, 9, 157

\end{thebibliography}

\end{document}